\documentclass[onecolumn]{article}

\usepackage[utf8]{inputenc}
\usepackage[margin=0.75in]{geometry}
\usepackage{authblk}
\usepackage{setspace}
\usepackage{graphicx}
\usepackage{tabularx}
\graphicspath{{./figures/}}
\usepackage{subcaption}
\usepackage{amsmath}
\usepackage{booktabs}
\usepackage{hyperref}
\usepackage{svg}

\usepackage[switch]{lineno}       
\usepackage[style=numeric, 
            sorting=none,        
            backref=false,
            backend=biber]{biblatex}

\title{ Multilayer Screening of Double and Conventional Perovskite Solar Cells Using SCAPS-1D and Machine Learning: Optimization of ETL, HTL, and Absorber for High-Efficiency Architectures}

\author[ a$\dag$]{ Neda Nasiri}
\author[ a$\dag$]{ Seyed Mahdi Mastoor}
\author[ a*]{ Amirhosein Ahmadkhan Kordbacheh}

\affil[ a]{ Department of Physics, Iran University of Science and Technology, Tehran, Iran}

\date{}

\begin{document}

\maketitle
{
\renewcommand{\thefootnote}{}
\footnotetext[0]{ $^\dagger$ These authors contributed equally to this work.}
\footnotetext{* Corresponding author: akordbacheh@iust.ac.ir}
}
\begin{abstract}
The combinatorial design space of multilayer perovskite solar cells is vast, yet exhaustive experimental or computational screening of all possible material combinations remains impractical. Here, we integrate SCAPS‑1D device simulations with machine learning to systematically explore 125 device architectures constructed from five electron transport layers (ETL), five absorbers (including lead‑free double perovskites), and five hole transport layers (HTL). A representative subset of configurations is used to train machine learning (ML) model, which predicts the power conversion efficiency (PCE) of the remaining unexplored structures. Leave‑One‑Group‑Out cross‑validation yields a Spearman rank correlation, demonstrating reliable ranking capability. SHAP (SHapley Additive exPlanations) analysis reveals that the HTL band gap, absorber band gap, and ETL electron affinity are the most influential descriptors, providing physical insights into interfacial recombination and charge extraction. The machine‑learning model identifies several high‑performance configurations that are subsequently verified by full SCAPS‑1D simulations. Among them, the device FTO/TiO$_2$/Cs$_2$AgBiBr$_6$/NiO/Ag achieves a PCE of 28.85$\%$, and the ML‑suggested structure FTO/SnO$_2$/Cs$_2$AgInBr$_6$/NiO/Ag exhibits 28.62$\%$, outperforming a closely related literature architecture by $\approx 4\%$ absolute. Notably, eight of the top‑11 structures employ the lead‑free double perovskite Cs$_2$AgInBr$_6$. This work demonstrates that a physics‑based, data‑driven workflow combining SCAPS‑1D, ML, and SHAP can accelerate the discovery of high‑efficiency, environmentally friendly perovskite solar cells while providing transparent design rules. The approach is generalizable to other multilayer optoelectronic systems.
\end{abstract}
\vspace{3em}


\section{Introduction}
Photovoltaic technology has gained significant attention in recent decades. Solar cells can reduce dependence on fossil fuels. Given the increasing global demand for electricity, this technology is an important area of research. Solar energy is widely available and has low environmental impact compared to fossil fuels \cite{Saad2025-we}.

Solar cells are categorized into three generations. Perovskite solar cells (PSCs) are a third‑generation photovoltaic technology. Their rapid efficiency progress has attracted significant attention. The optoelectronic properties of these materials enable high light absorption, allowing them to harvest a broad range of the solar spectrum\cite{Li2026-zm}. Due to their high compositional flexibility, many perovskites can be incorporated into different structures to investigate their behavior \cite{Kim2025-fp}. On the other hand, short-term stability remains one of the most significant drawbacks of PSCs \cite{Ahmed2025-yt}. This challenge has been addressed by another class of perovskites known as double perovskites (DPs). DPs, introduced with the $A_2BB'X_6$ structure, have demonstrated improved stability \cite{Biswas2026-zp}. DPs remove toxic lead from the B site and replace it with two less toxic
cations, which benefits environmental safety. This atomic engineering can also lead to more ordered and
symmetric crystal structures \cite{Rangar2025-ao}. Given the wide range of these materials, numerous structures can be
fabricated and investigated. Although much research has been conducted in this field, research gaps remain.

Sangavi et al.\cite{Sangavi2025-en} introduced Sm$_2$NiMnO$_6$ as a new lead-free material as a stable and less toxic alternativefor the absorber layer in perovskite solar cells. For the first time, they synthesized and tested this material. The group proposed using double perovskite materials instead of lead-halide perovskites. By removing Pb
and replacing it with Ni and Mn, they introduced a new fully inorganic material. The results for a
specific structure showed an efficiency of 9$\%$ through simulation and 4.3$\%$ through experiment. These
values were obtained for the structure selected by this team, which retained 86.6$\%$ of its initial current
density after 300 hours. Throughout the research conducted on this material, a gap is evident regarding
investigations with various structures. This material can be examined in numerous structures to identify the
most suitable one\cite{Hossain2026-xv}.

In a study, Fatmi et al. \cite{Fatmi2025-ec} demonstrated that Cs$_2$AgInBr$_6$ has a direct bandgap, is stable, and exhibits a high absorption coefficient, which allows for a thinner absorber layer while still achieving high efficiency
\cite{moses}. These findings highlight the material’s potential, though numerous structures incorporating it remain
unexplored. In contrast, Cs$_2$AgBiBr$_6$ faces the key challenge of a wide and indirect bandgap \cite{Borah2026-fq}; finding a suitable structure for this material could also lead to high efficiency, but existing studies have shown low
efficiencies \cite{Li2022-mb}\cite{Pang2022-py}\cite{Ou2022-cl}, and many compositions and structures are yet to be investigated.

Among the other materials used in this study are MAPbBr$_3$(CH$_3$NH$_3$PbBr$_3$) and MAPbI$_3$
(CH$_3$NH$_3$PbI$_3$), which are conventional perovskite materials. These materials have achieved high
efficiencies in previous studies\cite{Singh2026-pe}\cite{Boretti2026-dx}. However, despite their acceptable efficiency, there is still the possibility of achieving higher positions and efficiencies \cite{McGovern2021-ct}\cite{Chen2026-va}.

MASnBr$_3$ is a perovskite that has performed consistently well in previous studies and achieved high
efficiencies\cite{Mirdoraghi2025-fp}\cite{Hasan2025-te}. Recently, researchers have also used this material as a hole transport layer (HTL), yielding highly promising results\cite{Dakua2025-lx}\cite{Abrar2025-cw}. Due to the novelty of this idea, there are many structures in which this material has not yet been placed or investigated as the hole transport layer.

The remaining hole transport layers (HTLs) employed in this study include NiO \cite{Singh2026-nw}, CuO \cite{Lotfy2025-lc}, Spiro-OMeTAD \cite{Roula2025-tz}, and GaAs \cite{Khan2025-tu}. For the electron transport layer (ETL), ZnSe \cite{Monga2025-lj}, TiO$_2$ \cite{Shafique2026-fw}, PCBM \cite{Saidani2025-vr},
SnO$_2$\cite{Zhou2025-wb}, and CdZnS \cite{Talukder2025-uh} are utilized. These materials have demonstrated high efficiencies in prior investigations and are thus considered suitable candidates for integration with the other materials examined in the present research.

Although many of the ETL, absorber, and HTL materials considered in this study have been
individually investigated in previous works, the systematic exploration of their possible combinations remains limited. The large number of potential device architectures makes exhaustive simulation
computationally expensive and time-consuming. Consequently, many promising multilayer perovskite
configurations remain unexplored, highlighting the need for efficient screening strategies capable of
identifying high-performance structures within large compositional design spaces.

In recent years, machine learning (ML) has emerged as a powerful tool in physics and materials science,
enabling the discovery and optimization of complex systems through data-driven approaches \cite{Mastoor2026-jo}\cite{Peivaste2025-hj}. In the field of
photovoltaics, ML has been successfully applied to predict material properties\cite{Mao2025-oh}, optimize device parameters,
and accelerate the identification of high-performance solar-cell architectures\cite{Shafian2025-xs}. By reducing the need for exhaustive simulations and experiments, ML offers an efficient framework for exploring large design
spaces and guiding the development of next-generation photovoltaic devices.

In this work, a combinatorial design space consisting of 125 multilayer perovskite solar-cell configurations was constructed using five ETL, five absorber, and five HTL materials, including both conventional and double-perovskite absorbers. Fluorine-doped tin oxide (FTO) and silver (Ag) were employed as the front and back contacts, respectively. SCAPS-1D simulations and machine learning were employed as complementary tools to efficiently explore this design space. The proposed framework enabled the screening of previously unexplored device architectures and the identification of several high-performance configurations with photovoltaic efficiencies exceeding those of many initially investigated structures. Furthermore, SHAP (SHapley Additive exPlanations) analysis was used to provide physical insight into the material properties governing device performance.

The remainder of this paper is organized as follows; Section 2 describes the simulation parameters,
dataset construction, and machine learning methodology. Section 3 presents the results, including model
validation, SHAP analysis, and the discovered high-efficiency structures, along with a discussion of their
physical implications. Section 4 concludes the paper.

\section{Methods}
\begin{figure}
    \centering
    \includegraphics[width=0.5\linewidth]{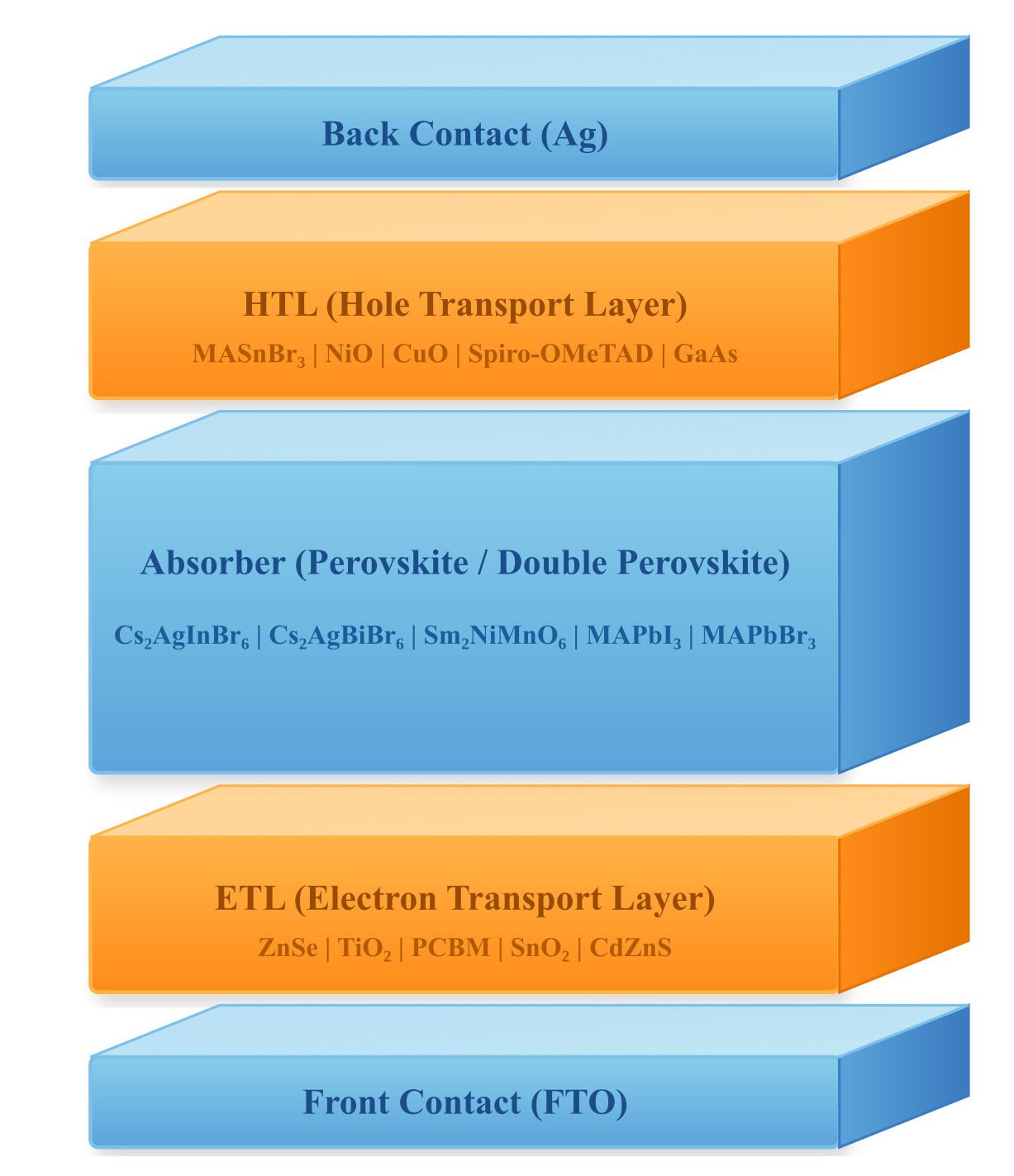}
    \caption{Schematic illustration of the multilayer perovskite solar-cell architecture used in this study, showing the front contact/ETL/absorber/HTL/back contact configuration and the materials considered for each layer.}
    \label{fig:structture}
\end{figure}

The perovskite solar cell structure adopted in this study follows the configuration: Front contact / ETL / Perovskite absorber / HTL / Back contact, Each layer was designated with specific materials. Ag
and FTO served as the back contact and front contact, respectively, with work function of 4.4 eV \cite{Abrar2025-cw}and 4.7 eV \cite{Taheri2021-nr}. As
shown in fig.\ref{fig:structture} For the ETL, absorber layer and HTL, we selected five materials for each.

The material set was selected from candidates previously demonstrated in the literature to perform well as ETL, absorber, or HTL materials in solar-cell architectures. The selection was guided by a combination of reported photovoltaic efficiency, band-edge alignment, carrier-transport characteristics, and compatibility with multilayer device operation, rather than by standalone material performance alone. It should be noted that interfacial mismatch between layers, even with
individually high-performing materials, can lead to $V_{oc}$ losses and consequently a dramatic drop in PCE. The subset of these structures was simulated using the SCAPS software.

SCAPS simulator is a solar cell simulation software. The version used in this study is the one-dimensional version,
which was developed by Marc Burgelman at Ghent University in Belgium. After defining the layer parameters and
simulation conditions, SCAPS-1D calculates the device performance by solving the Poisson equation and the coupled continuity equations for electrons and holes.
\begin{equation}
-\frac{\partial}{\partial x} \left( -\epsilon(x) \frac{\partial V}{\partial x} \right) = q \left[ p(x) - n(x) + N_D^+(x) - N_A^-(x) + p_t(x) - n_t(x) \right]
\label{eq:poisson}
\end{equation}
\begin{equation}
\frac{\partial p}{\partial t} = \frac{1}{q} \frac{\partial J_p}{\partial x} + G_p - R_p
\label{eq:continuity_holes}
\end{equation}
\begin{equation}
\frac{\partial n}{\partial t} = \frac{1}{q} \frac{\partial J_n}{\partial x} + G_n - R_n
\label{eq:continuity_electrons}
\end{equation}
In the Poisson equation (Equation~\ref{eq:poisson}), $x$ is the spatial coordinate, $\epsilon(x)$ is the position-dependent dielectric permittivity, $V$ is the electrostatic potential, $q$ is the elementary charge, $p(x)$ is the free hole concentration, $n(x)$ is the free electron concentration, $N_D^+(x)$ is the ionized donor concentration, $N_A^-(x)$ is the ionized acceptor concentration, $p_t(x)$ is the trapped hole concentration, and $n_t(x)$ is the trapped electron concentration. In the continuity equation for holes (Equation~\ref{eq:continuity_holes}), $p$ is the hole concentration, $t$ is time, $J_p$ is the hole current density, $G_p$ is the hole generation rate, and $R_p$ is the hole recombination rate. In the continuity equation for electrons (Equation~\ref{eq:continuity_electrons}), $n$ is the electron concentration, $t$ is time, $J_n$ is the electron current density, $G_n$ is the electron generation rate, and $R_n$ is the electron recombination rate.

Simulations are performed using the electrical and optical properties listed in Tables \ref{tab:ETL}, \ref{tab:HTL}, and \ref{tab:absorber}, in which these
parameters describe the properties of each material.

\begin{table}
    \centering
    \begin{tabular}{|l|c|c|c|c|c|}\hline
         parameters&  ZnSe&  TiO$_2$&  PCBM&  SnO$_2$& CdZnS\\\hline
         Thinckness (nm)&  100&  100&  100&  100& 100\\\hline
         electron affinity (eV)&  4.09&  4.1&  4.2&  4& 4.2\\\hline
         band gap (eV)&  2.81&  3.2&  2&  3.5& 3.2\\\hline
         dielectric permittivity&  8.6&  9&  3.9&  9& 9.12\\\hline
         CB effective density of state (cm$^{-3}$)&  $2.2 \times 10^{18}$&  $2.2 \times 10^{18}$&  $2.5 \times 10^{21}$&  $2.2 \times 10^{17}$& $1.5 \times 10^{18}$\\\hline
         VB effective density of state (cm$^{-3}$)&  $1.8 \times 10^{18}$&  $1 \times 10^{19}$&  $2. \times 10^{21}$&  $2.2 \times 10^{16}$& $1.8 \times 10^{18}$\\\hline
         electron mobility (cm$^2$/Vs)&  $5 \times 10^{2}$&  20&  0.2&  240& 250\\\hline
         hole mobility (cm$^{2}$/Vs)&  $1.1 \times 10^{2}$&  10&  0.2&  220& 40\\\hline
 shallow uniform donor density ND (cm$^{-3}$)& $1 \times 10^{15}$& $1 \times 10^{18}$& $1 \times 10^{18}$& $2.42 \times 10^{19}$&$1 \times 10^{16}$\\\hline
 shallow uniform acceptor density NA (cm$^{-3}$)& 0& 0& 0& 0&0\\\hline
 electron thermal velocity (cm/s)& $1 \times 10^{7}$& $1 \times 10^{7}$& $1 \times 10^{7}$& $1 \times 10^{7}$&$1 \times 10^{7}$\\\hline
         hole thermal velocity (cm/s)&  $1 \times 10^{7}$&  $1 \times 10^{7}$&  $1 \times 10^{7}$&  $1 \times 10^{7}$& $1 \times 10^{7}$\\\hline
 ref& \cite{Abrar2025-cw}& \cite{Alkhammash2023-uu}&\cite{Alkhammash2023-uu} &\cite{Alkhammash2023-uu} &\cite{Uddin2024-cb}\\ \hline
    \end{tabular}
    \caption{Input parametrs for ETL}
    \label{tab:ETL}
\end{table}

\begin{table}[]
    \centering
    \begin{tabular}{|l|c|c|c|c|c|}\hline
         parameters&  MASnBr$_3$&  NiO&  CuO&  Spiro-OMeTAD& GaAs\\\hline
         Thinckness (nm)&  200&  200&  200&  200& 200\\\hline
         electron affinity (eV)&  3.39&  1.8&  4.07&  2.05& 4.07\\\hline
         band gap (eV)&  2.15&  3.6&  1.48&  2.8& 1.42\\\hline
         dielectric permittivity&  8.2&  11.7&  18.1&  3& 12.9\\\hline
         CB effective density of state (cm$^{-3}$)&  $1 \times 10^{20}$&  $2.5 \times 10^{20}$&  $2.1 \times 10^{19}$&  $2.2 \times 10^{18}$& $2.2 \times 10^{18}$\\\hline
         VB effective density of state (cm$^{-3}$)&  $1 \times 10^{18}$&  $2.5 \times 10^{20}$&  $2.5 \times 10^{19}$&  $1.8 \times 10^{19}$& $1.8 \times 10^{19}$\\\hline
         electron mobility (cm$^2$/Vs)&  1.6&  2.8&  100&  0.0002& 850\\\hline
         hole mobility (cm$^{2}$/Vs)&  2&  2.8&  0.1&  0.0002& 400\\\hline
 shallow uniform donor density ND (cm$^{-3}$)& 0& 0& 0& 0&0\\\hline
 shallow uniform acceptor density NA (cm$^{-3}$)& $1 \times 10^{18}$& $3 \times 10^{18}$& $1 \times 10^{18}$& $1 \times 10^{18}$&$1 \times 10^{17}$\\\hline
 electron thermal velocity (cm/s)& $1 \times 10^{7}$& $1 \times 10^{7}$& $1 \times 10^{7}$& $1 \times 10^{7}$&$1 \times 10^{7}$\\\hline
         hole thermal velocity (cm/s)&  $1 \times 10^{7}$&  $1 \times 10^{7}$&  $1 \times 10^{7}$&  $1 \times 10^{7}$& $1 \times 10^{7}$\\\hline
 ref& \cite{Abrar2025-cw}& \cite{Alkhammash2023-uu}&\cite{Alkhammash2023-uu} &\cite{Alkhammash2023-uu} &\cite{Uddin2024-cb}\\ \hline
    \end{tabular}
    \caption{Input parametrs for HTL}
    \label{tab:HTL}
\end{table}

\begin{table}
    \centering
    \begin{tabular}{|l|c|c|c|c|c|}\hline
         parameters&  Cs$_2$AgInBr$_6$&  MAPbI$_3$&  Cs$_2$AgBiBr$_6$&  Sm$_2$NiMnO$_6$& MAPbBr$_3$\\\hline
         Thinckness (nm)&  600&  600&  600&  600& 600\\\hline
         electron affinity (eV)&  4.1&  3.93&  4.19&  3.52& 3.7\\\hline
         band gap (eV)&  1.47&  1.5&  1.7&  1.5& 2.2\\\hline
         dielectric permittivity&  4.37&  30&  5.8&  3.5& 10\\\hline
         CB effective density of state (cm$^{-3}$)&  $1.26\times 10^{18}$&  $2.5 \times 10^{20}$&  $1\times 10^{19}$&  $1\times 10^{18}$& $1\times 10^{17}$\\\hline
         VB effective density of state ($cm^{-3}$)&  $1.73 \times 10^{18}$&  $2.5 \times 10^{20}$&  $1\times 10^{19}$&  $1 \times 10^{18}$& $1\times 10^{18}$\\\hline
         electron mobility (cm$^2$/Vs)&  89.4&  50&  11.81&  22& 24\\\hline
         hole mobility (cm$^{2}$/Vs)&  3.3&  50&  0.49&  22& 24\\\hline
 shallow uniform donor density ND (cm$^{-3}$)& 0& 0& $1\times 10^{16}$& $1\times 10^{15}$&$1\times 10^{13}$\\\hline
 shallow uniform acceptor density NA (cm$^{-3}$)& $1 \times 10^{18}$& $3 \times 10^{18}$& $1 \times 10^{18}$& $1 \times 10^{18}$&$1 \times 10^{17}$\\\hline
 electron thermal velocity (cm/s)& $1 \times 10^{7}$& $1 \times 10^{7}$& $1 \times 10^{7}$& $1 \times 10^{7}$&$1 \times 10^{7}$\\\hline
         hole thermal velocity (cm/s)&  $1 \times 10^{7}$&  $1 \times 10^{7}$&  $1 \times 10^{7}$&  $1 \times 10^{7}$& $1 \times 10^{7}$\\\hline
 ref&\cite{Abrar2025-cw} &\cite{Taheri2021-nr} & \cite{Alkhammash2023-uu}&\cite{Sangavi2025-en} &\cite{Fahim2026-fw}\\ \hline
    \end{tabular}
    \caption{Input parametrs for absorber}
    \label{tab:absorber}
\end{table}

Accurate modeling is achieved only when these two property categories are coupled. Crucially, it is the synergy
between them, not their individual effects, that determines the accuracy of the model. After completing the
simulations, the outputs are obtained in the form of graphs and numerical parameters.
Our main focus in this project was on the J-V. From the J-V curve, the open-circuit voltage (V$_{oc}$),
short-circuit current (J$_{sc}$), fill factor (FF) and Power Conversion Efficiencies (PCE) parameters can be extracted.
The simulation conditions must comply with standard test conditions (STC). This means a standard temperature of 300 K, which corresponds to room temperature. The reason for this is to enable researchers to replicate the experiment in the laboratory.
The
prerequisite for PCE calculation is that the simulation must be performed under illumination, not in dark mode, and
an appropriate light spectrum must be selected. This spectrum global standard is AM 1.5G which accounts for both direct and diffuse sunlight affecting the device performance\cite{Sevillano-Bendezu2023-kv}.

\subsection{Data construction and Machine learning}

The dataset was constructed from a total design space of 125 possible configurations, from which a representative subset of 37 structures (approximately $30 \%$  of the entire set) was selected. The selection was not performed randomly; rather, it was guided by the objective of maximizing diversity across material combinations while minimizing redundancy. In particular, it was ensured that the chosen configurations span a wide range of structural and compositional variations, thereby reducing the risk of overfitting during model training. Additionally, care was taken to guaranty that each constituent material appears multiple times within the dataset, allowing its contribution to the target properties to be adequately learned by the model. This approach resulted in a balanced and informative dataset that supports robust and generalizable machine learning performance.

A set of physically relevant material descriptors was employed as input features, including band gap, electron affinity, dielectric permittivity, conduction band effective density of states, valence band effective density of states, electron mobility, hole mobility, shallow uniform donor density (ND), shallow uniform acceptor density (NA), and absorption constant (A). These features were selected based on their direct influence on the fundamental physical processes governing solar cell performance. Specifically, parameters such as band gap and absorption constant determine the light-harvesting capability \cite{Chen2017-ml}\cite{Zdanowicz2005-yo}, While electron affinity and dielectric permittivity affect band alignment and charge separation, respectively, by controlling interfacial energy offsets and Coulomb screening\cite{Hood2016-ee}\cite{Klein2012-fz}. Carrier transport properties, including electron and hole mobilities as well as effective densities of states, play a critical role in charge transport and recombination dynamics\cite{Musiienko2024-nq}. Furthermore, doping concentrations (ND and NA) control carrier concentration and internal electric fields within the device \cite{Movla2023-tv}.
In addition, the categorical information associated with material configurations was encoded using one-hot encoding, through which the presence or absence of each material component within a given configuration was represented in a binary format. This combined feature representation enabled both intrinsic material properties and compositional information to be systematically incorporated into the model.

In this work, the relationship between device configuration and PCE which represent by ($\eta$) was modeled using the XGBoost regression algorithm. This choice was guided by both the nature of the dataset and the requirements of the problem to sort $\eta$. The available data are relatively limited in size and consist of structured inputs that combine discrete material configurations with physically meaningful descriptors. Under such conditions, tree-based ensemble methods have been shown to perform reliably without the need for large training samples\cite{Tsiligaridis2023-iq}\cite{Jiang2025-ne}.
Another important consideration is the inherently nonlinear behavior of multilayer perovskite solar cells, where the contribution of each layer is strongly coupled to the others. XGBoost is well suited to capture these interactions through its boosting mechanism, which incrementally refines predictions by focusing on previously misestimated patterns \cite{Ahamed2026-jp}\cite{Bibi2025-kb}. In addition, the method includes regularization and implicit feature selection, which help to control model complexity and reduce the likelihood of overfitting, particularly in settings with multiple correlated inputs \cite{De_la_Asuncion-Nadal2025-ze}. This ML model also offers a good balance between predictive performance and interpretability. Compared to more complex approaches such as deep neural networks, it provides more stable training behavior and allows clearer insight into feature importance \cite{Shwartz-Ziv2022-dy}, which is valuable for understanding the physical factors governing the efficiency of devices. 

Each device configuration was assigned a unique device identifier, and all samples associated with the same device were kept within a single fold during validation. For robust evaluation of generalization performance, Leave-One-Group-Out Cross Validation (LOGO-CV) was employed, in which one device configuration was excluded during each iteration and used exclusively for testing, while the remaining configurations were used for training.
The predictive performance of the model was quantified using the root mean square error (RMSE), mean absolute error (MAE), coefficient of determination ($R^2$), and Spearman rank correlation coefficient. RMSE and MAE were calculated as
\begin{equation}
    \text{RMSE} = \sqrt{\frac{1}{n} \sum_{i=1}^{n} (y_i - \hat{y}_i)^2} 
\end{equation}
\begin{equation}
\text{MAE} = \frac{1}{n} \sum_{i=1}^{n} |y_i - \hat{y}_i|
\end{equation}
where $y_i$and $\hat{y}_i$ are the actual and predicted efficiencies, respectively. 
The coefficient of determination was computed according to:
\begin{equation}
    R^2 = 1 - \frac{\sum_{i=1}^{n} (y_i - \hat{y}_i)^2}{\sum_{i=1}^{n} (y_i - \bar{y})^2}
\end{equation}
where $\bar{y}$ denotes the mean of the observed efficiencies. In addition, the Spearman rank correlation coefficient was used to evaluate the ranking capability of the model for screening applications.
\begin{equation}
    \rho = 1 - \frac{6 \sum d_i^2}{n(n^2 - 1)}
\end{equation}
where $d_i$ represents the difference between the predicted and actual ranks of the $i-th$ configuration.

\begin{figure}[]
    \centering
    \includegraphics[width=1\linewidth]{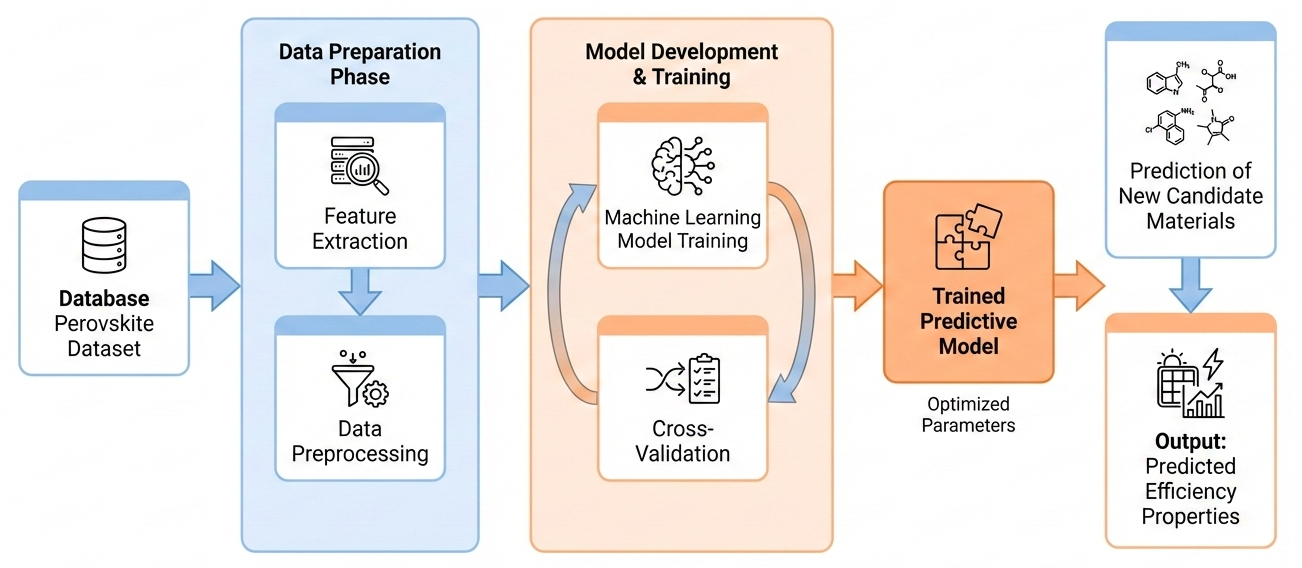}
    \caption{Workflow of the proposed SCAPS-1D and machine-learning framework, showing dataset construction from the simulated device configurations, model training and validation, screening of unexplored structures, and verification of the top-ranked candidates with full SCAPS-1D simulations.}
    \label{workflow}
\end{figure}

After model validation, the trained XGBoost model was employed to predict the efficiencies of unexplored perovskite solar-cell configurations within the remaining combinatorial design space. Candidate structures with the highest predicted efficiencies were subsequently selected for further verification using full SCAPS-1D simulations. This workflow as clarified in fig.\ref{workflow} facilitated a faster exploration of the configuration space while significantly reducing the computational cost associated with exhaustive device simulation.

\section{Results }
The PCE ($\eta$) values exhibited a broad distribution, ranging from approximately $ 1\%$ to $28.5\%$, with an average efficiency of about $14.8\%$. This wide variation in device performance indicates that the dataset contains both low and high efficiency configurations, providing sufficient diversity for the machine-learning model to learn meaningful relationships between material properties, device architecture, and photovoltaic performance. Furthermore, all constituent materials considered within the complete design space were represented within the training dataset, ensuring that the model was exposed to the full range of material types during training.

\begin{figure}[]
    \centering
    \includegraphics[width=0.7\linewidth]{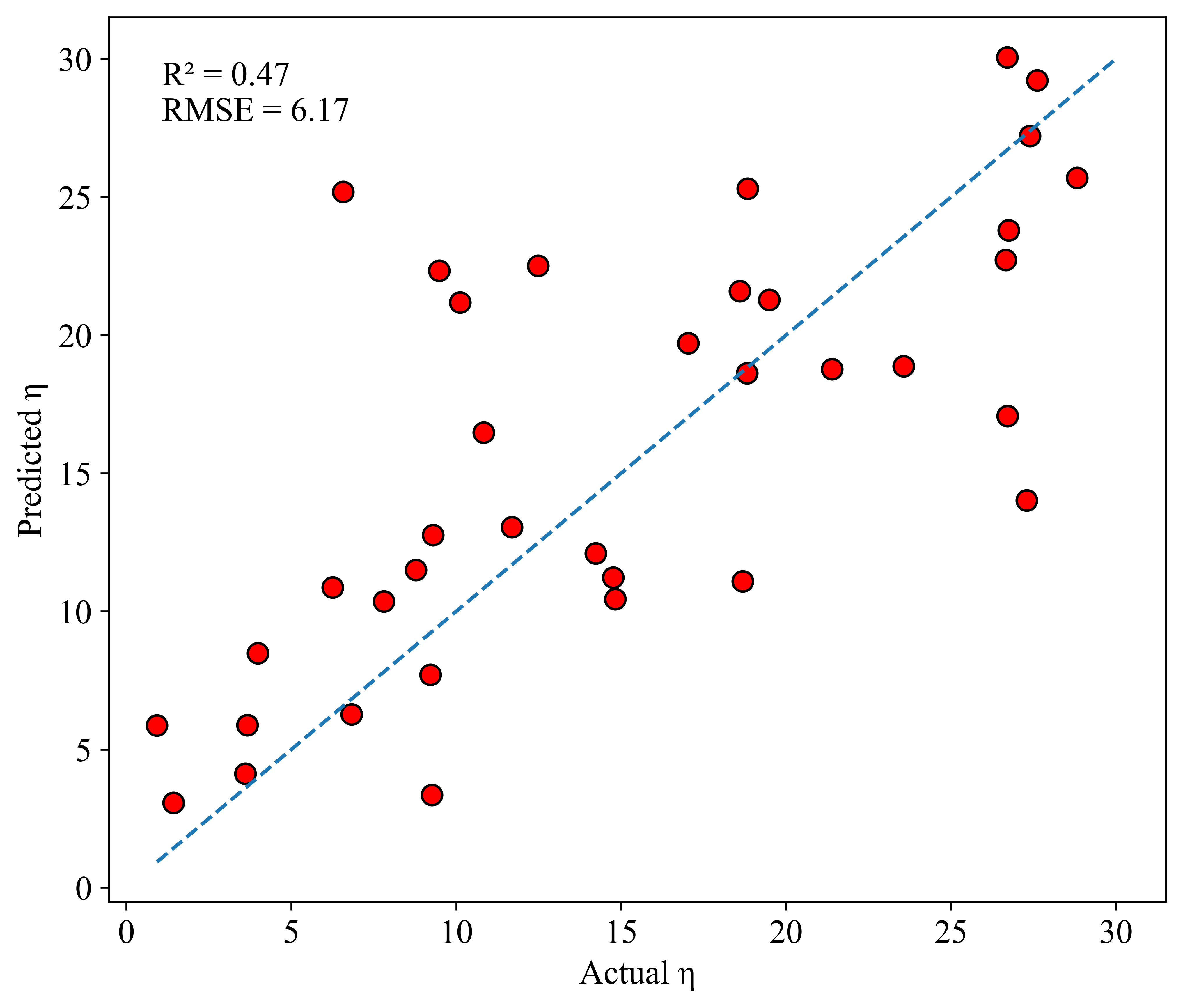}
    \caption{Predicted versus actual power conversion efficiency (PCE) for the out-of-sample Leave-One-Group-Out Cross-Validation (LOGO-CV) predictions. Each point represents a device configuration. The diagonal dashed line indicates ideal agreement between predicted and simulated efficiencies.}
    \label{fig:scater}
\end{figure}

The predictive performance of the developed XGBoost model was evaluated using Leave-One-Group-Out Cross Validation (LOGO-CV), in which each device configuration was treated as an independent group and excluded entirely during testing. The obtained validation results demonstrate that the model is capable of capturing the general relationship between material configuration and photovoltaic efficiency with reasonable reliability. The model achieved an overall RMSE of 6.17 and an MAE of 4.64, while the coefficient of determination reached $R^2=0.47$. In addition, the Spearman rank correlation coefficient was calculated as 0.74, indicating that the model preserves the relative ranking of device efficiencies with good consistency. Figure \ref{fig:scater} presents the predicted-versus-actual efficiency scatter plot obtained from the out-of-sample LOGO-CV predictions. A clear positive correlation between predicted and simulated efficiencies can be observed, indicating that the model successfully captures the overall efficiency trend across different device configurations. Although several configurations deviate from the ideal diagonal line, particularly in the high-error region, the general agreement between predicted and actual values confirms the capability of the model to identify promising high-efficiency structures. The observed deviations are likely associated with the complex nonlinear interactions among multilayer material combinations. Overall, the obtained results suggest that the developed machine-learning framework is reasonably reliable for screening and prioritizing unexplored perovskite solar-cell configurations, even though its absolute prediction accuracy remains moderate.

\begin{figure}
    \centering
    \includegraphics[width=0.7\linewidth]{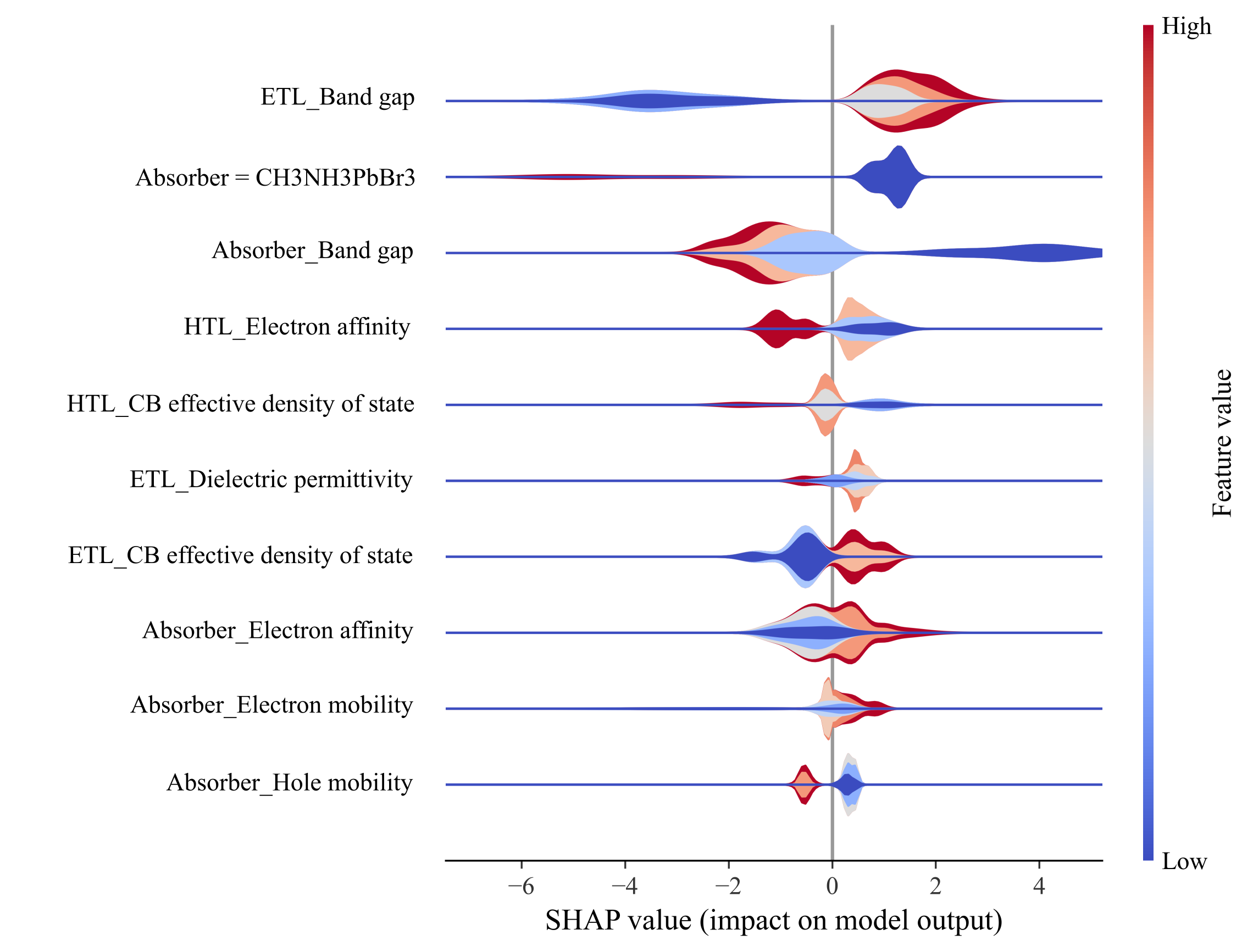}
    \caption{SHAP summary plot showing the global feature importance ranking for PCE. Features are ranked by importance, with color indicating feature value (red: high, blue: low).}
    \label{fig:shap}
\end{figure}

To elucidate the global feature importance and decode the underlying optoelectronic mechanisms governing the PCE of the multi-layer photovoltaic cell, a SHAP summary analysis was executed (Figure \ref{fig:shap}).
The resulting hierarchy demonstrates that the energetic configurations of the layers dominate device performance, led by the band gap of the ETL. Its distinct bimodal distribution reveals that a wider band gap in this layer significantly bolsters PCE (positive SHAP values up to +3), acting as an effective carrier-blocking or window layer that suppresses non-radiative interfacial recombination. Conversely, an inverse trend is resolved for the primary absorber layer, where a lower band gap yields a pronounced positive impact extending to +5 SHAP values, structurally aligning with standard photovoltaic principles where a narrower gap extends photon harvesting into the near-infrared spectrum to maximize short-circuit current \cite{Miah2024-rk}. Crucially, the model captures the detrimental impact of incorporating wide-bandgap materials within the absorber matrix, the explicit presence of $\text{CH}_3\text{NH}_3\text{PbBr}_3$ in absorber layer (represented by the red feature values extending into the negative SHAP region) severely penalizes efficiency, whereas its absence (blue bulb clustered at +1.2) consistently yields superior performance by allowing optimal, narrower-gap alternatives. Beyond macroscopic energy gaps, the SHAP plot successfully maps localized transport and extraction dynamics. Efficient charge collection is driven by a lower electron affinity in the front layer and a higher conduction band effective density of states in the rear layer, both of which minimize interfacial extraction barriers and facilitate rapid carrier sweep-out. At the lower end of the hierarchy, the model identifies subtle minority carrier effects; for instance, a lower hole mobility in absorber layer (Absorber\_Hole mobility , indicated by the negative SHAP impact of red feature values) is marginally favored, pointing to a potential mitigation of unwanted carrier back-transfer or leakage currents. Ultimately, this data-driven analysis maps a rigorous, physically intuitive architectural blueprint: optimizing multi-layer perovskite PCE demands a narrow bandgap active absorber paired with highly asymmetric, wide bandgap transport layers engineered with high effective densities of states to secure rapid, unidirectional charge extraction.

The trained framework was used to estimate the PCE of the remaining 88 unexplored device configurations in the full combinatorial design space. Based on the predicted efficiencies, the 10 candidate structures with the highest estimated PCE values were selected for further verification using full SCAPS-1D simulations. The SCAPS-1D calculations confirmed that several of the ML-selected configurations exhibited high photovoltaic performance, demonstrating that the proposed framework can effectively screen and prioritize promising device architectures. To further analyze the characteristics of high-efficiency structures, the validated ML-selected configurations were combined with the original training dataset, resulting in a total of 47 investigated devices. Among these, configurations with PCE values above 25$\%$ were identified and compared. Notably, several of the high-performance structures listed in Table \ref{tab:top11} were not part of the initial training dataset and were discovered exclusively through the machine-learning screening procedure. These results indicate that the ML-guided workflow can accelerate the identification of promising multilayer perovskite solar-cell configurations while significantly reducing the number of expensive SCAPS simulations required for exhaustive exploration of the full design space.

Among the structures identified by machine learning, FTO/SnO$_2$/Cs$_2$AgInBr$_6$/NiO/Ag achieved an efficiency of 28.62$\%$. This result is notable because a closely related architecture, FTO/SnO$_2$/Cs$_2$AgInBr$_6$/CuO/Ag, was previously reported by Bechane et al.\cite{Bechane2025-sr} to reach a lower PCE of 24.96$\%$ with V$_{OC}$=0.969\,V, J$_{sc}$=29.4\,mA/cm$^2$ and FF = 87.6$\%$. The replacement of NiO with CuO resulted in approximately a 4$\%$ efficiency difference, attributed to several factors. NiO exhibits faster charge transfer than CuO, which reduces recombination, and has a wider bandgap than CuO (as shown in Table \ref{tab:HTL}). This superior performance of the NiO-based structure is consistent with SHAP trends, which show that HTL-related descriptors, particularly band gap and charge-transport parameters, strongly influence device efficiency. Thus, the ML-guided screening successfully identified a more favorable transport-layer combination within the same absorber-based device family, supporting both the physical interpretation of the SHAP analysis and the practical value of the proposed workflow.

\begin{table}[]
	\centering
	\begin{tabular}{|l|c|c|c|c|}
		\hline
		Device structure                             & V$_{oc}$ & J$_{sc}$ & FF    & PCE   \\ \hline
		FTO/TiO$_2$/Cs$_2$AgBiBr$_6$/NiO/Ag          & 1.23     & 27.74    & 79.1  & 28.85 \\ \hline
		FTO/TiO$_2$/Cs$_2$AgInBr$_6$/NiO/Ag          & 1.23     & 27.77    & 78.95 & 28.82 \\ \hline
		* FTO/SnO$_2$/Cs$_2$AgInBr$_6$/NiO/Ag        & 1.23     & 27.76    & 78.35 & 28.62 \\ \hline
		FTO/ZnSe/Cs$_2$AgInBr$_6$/NiO/Ag             & 1.1      & 27.78    & 84.41 & 27.61 \\ \hline
		FTO/CdZnS/Cs$_2$AgInBr$_6$/NiO/Ag            & 1.1      & 27.65    & 84.14 & 27.39 \\ \hline
		FTO/ZnSe/CH$_3$NH$_3$PbI$_3$/NiO/Ag          & 1.22     & 26.80    & 77.90 & 27.29 \\ \hline
		* FTO/SnO$_2$/Cs$_2$AgInBr$_6$/MASnBr$_3$/Ag & 1.22     & 27.76    & 75.26 & 27.24 \\ \hline
		* FTO/ZnSe/Cs$_2$AgInBr$_6$/MASnBr$_3$/Ag    & 1.22     & 27.77    & 74.16 & 26.75 \\ \hline
		FTO/TiO$_2$/Cs$_2$AgBiBr$_6$/MASnBr$_3$/Ag   & 1.22     & 27.74    & 74.16 & 26.71 \\ \hline
		FTO/TiO$_2$/Cs$_2$AgInBr$_6$/MASnBr$_3$/Ag   & 1.22     & 27.74    & 74.13 & 26.70 \\ \hline
		FTO/PCBM/Cs$_2$AgInBr$_6$/NiO/Ag             & 1.1      & 27.32    & 82.96 & 26.66 \\ \hline
	\end{tabular}
	\caption{Photovoltaic performance parameters (open‑circuit voltage (V$_{OC}$) in V, short‑circuit current density (J$_{SC}$) in mA/cm$^2$, fill factor FF, and power conversion efficiency (PCE)) in \% of the top-performing perovskite solar cell configurations identified by machine learning and validated with SCAPS-1D simulations. Structures marked with an asterisk (*) were suggested by the machine learning model.}
	\label{tab:top11}
\end{table}

As observed, among the 11 top-performing structures, the absorber layer of one structure consists of a conventional perovskite, namely CH$_3$NH$_3$PbI$_3$, which achieved a PCE of 27.29$\%$, along with a FF of 77.9$\%$. The remaining 10 structures employ double perovskites, with 8 structures featuring a Cs$_2$AgInBr$_6$ absorber layer and 2 structures featuring a Cs$_2$AgBiBr$_6$ absorber layer. As presented in Table \ref{tab:top11}, the best result among the reported structures belongs to the device incorporating a Cs$_2$AgInBr$_6$ absorber layer, which achieved a PCE of 28.85$\%$ and an FF of 79.1$\%$. This structure employs NiO and TiO$_2$ as the HTL and ETL, respectively. It can be asserted that each of these materials exhibited the best performance among the candidate materials in their respective roles within this study.
\begin{figure}
    \centering
    \includegraphics[width=1\linewidth]{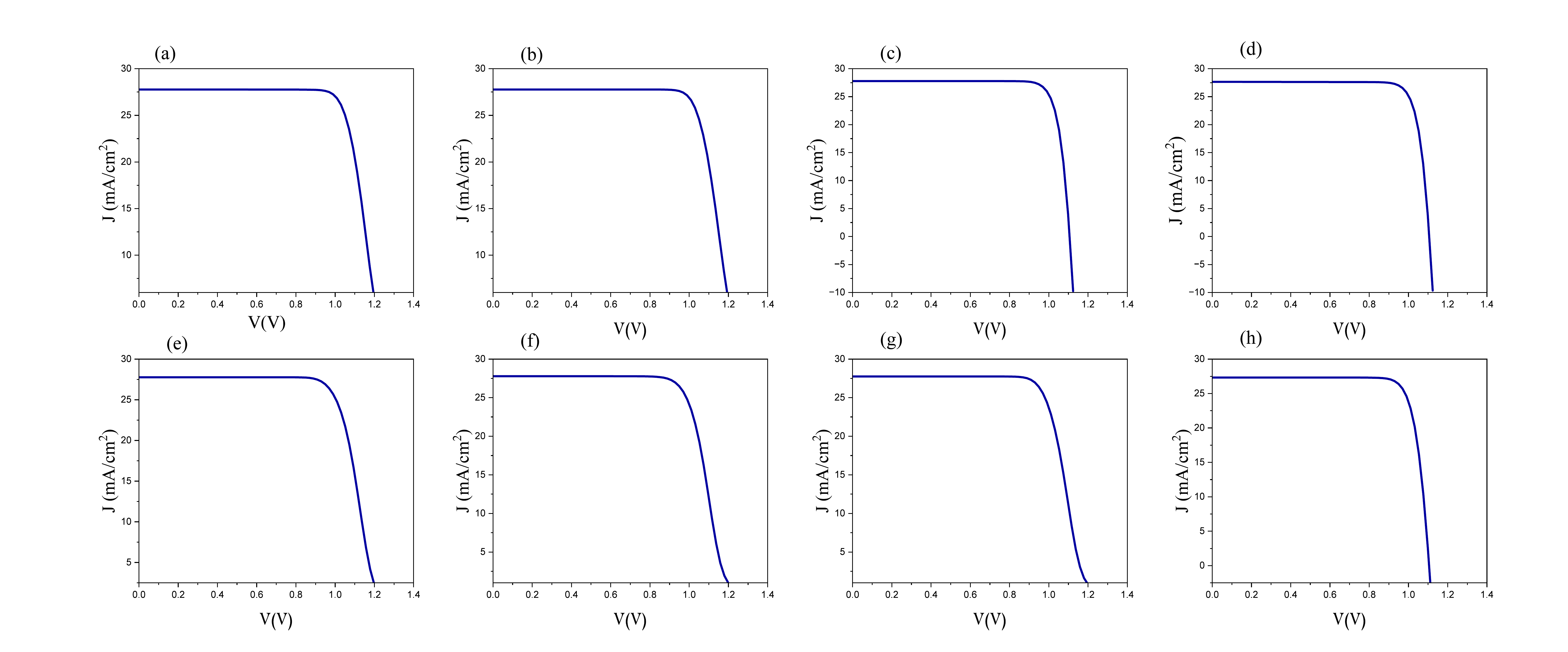}
    \caption{Current density–voltage (J‑V) characteristics of Cs$_2$AgInBr$_6$ based perovskite solar cells for the following configurations: (a) FTO/TiO$_2$/Cs$_2$AgInBr$_6$/NiO/Ag, (b) FTO/SnO$_2$/Cs$_2$AgInBr$_6$/NiO/Ag, (c) FTO/ZnSe/Cs$_2$AgInBr$_6$/NiO/Ag, (d) FTO/CdZnS/Cs$_2$AgInBr$_6$/NiO/Ag, (e) FTO/SnO$_2$/Cs$_2$AgInBr$_6$/MASnBr$_3$/Ag, (f) FTO/ZnSe/Cs$_2$AgInBr$_6$/MASnBr$_3$/Ag, (g) FTO/TiO$_2$/Cs$_2$AgInBr$_6$/MASnBr$_3$/Ag, and (h) FTO/PCBM/Cs$_2$AgInBr$_6$/NiO/Ag. Corresponding photovoltaic parameters are summarized in Table \ref{tab:top11}.}
    \label{fig:8}
\end{figure}

Fig \ref{fig:8} presents the J-V curves for the structures with a Cs$_2$AgInBr$_6$ absorber layer. Unlike other double perovskites, this material has a direct bandgap, which enhances light absorption and gives it superior performance as an absorber layer \cite{Han2026-yd}. As shown in Table \ref{tab:absorber}, due to its high electron mobility, after generation, electrons migrate rapidly toward the ETL. This correlates with the low effective mass of electrons, which are substantially lighter than holes, allowing them to traverse the absorber layer at high velocity and efficiently reach the ETL. These properties collectively reduce recombination, leading to an enhanced PCE. Additionally, Cs$_2$AgInBr$_6$ is a completely inorganic, lead‑free material, offering improved stability and environmental friendliness compared to conventional lead‑based perovskites.

\begin{figure}
    \centering
    \includegraphics[width=0.9\linewidth]{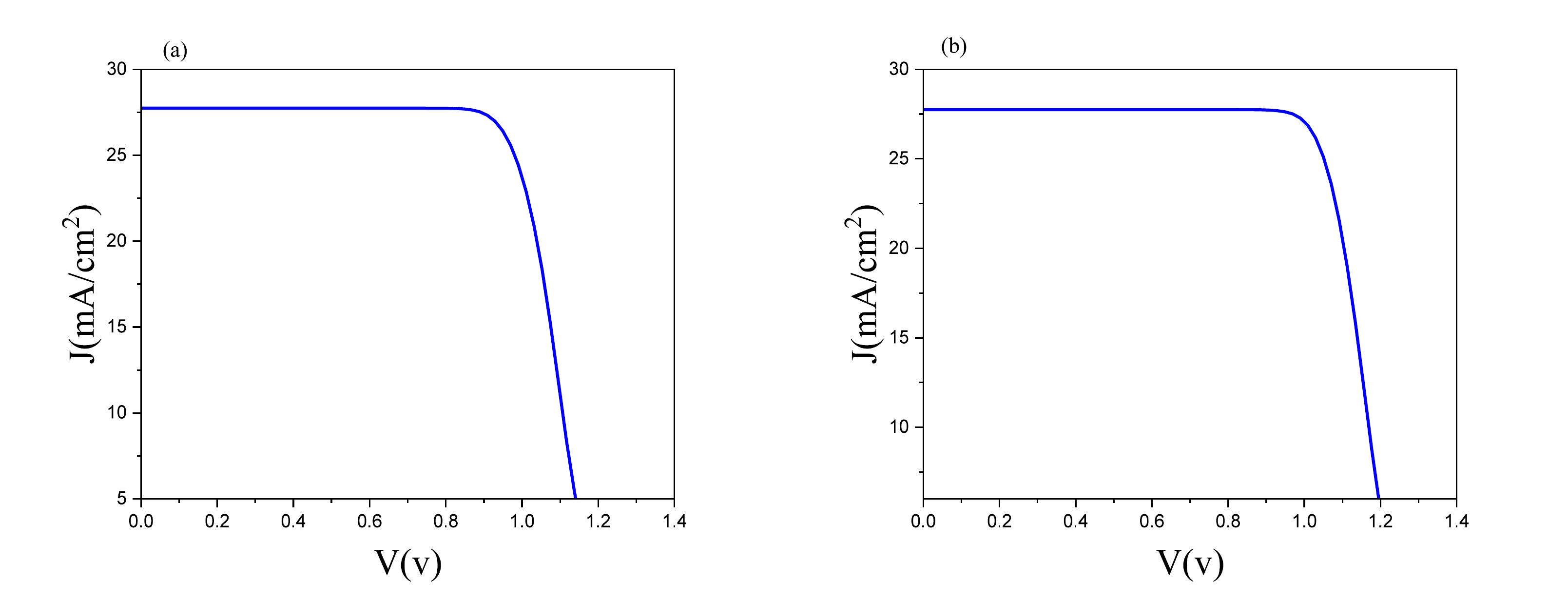}
    \caption{Current density–voltage (J‑V) characteristics of perovskite solar cells with the configurations (a) FTO/TiO$_2$/Cs$_2$AgBiBr$_6$/MASnBr$_3$/Ag and (b) FTO/TiO$_2$/Cs$_2$AgBiBr$_6$/NiO/Ag. Corresponding photovoltaic parameters are summarized in Table \ref{tab:top11}.}
    \label{fig:2}
\end{figure}

The results obtained from the structures featuring a Cs$_2$AgBiBr$_6$ absorber layer are presented in Fig \ref{fig:2} Although this material belongs to the similar family as the previously discussed compound, it exhibited inferior performance. This behavior can be attributed to its indirect bandgap \cite{Gupta2026-zp}. This characteristic results in less efficient light absorption compared to Cs$_2$AgInBr$_6$. Nevertheless, it still demonstrates superior performance relative to conventional perovskites.

\begin{figure}
    \centering
    \includegraphics[width=0.5\linewidth]{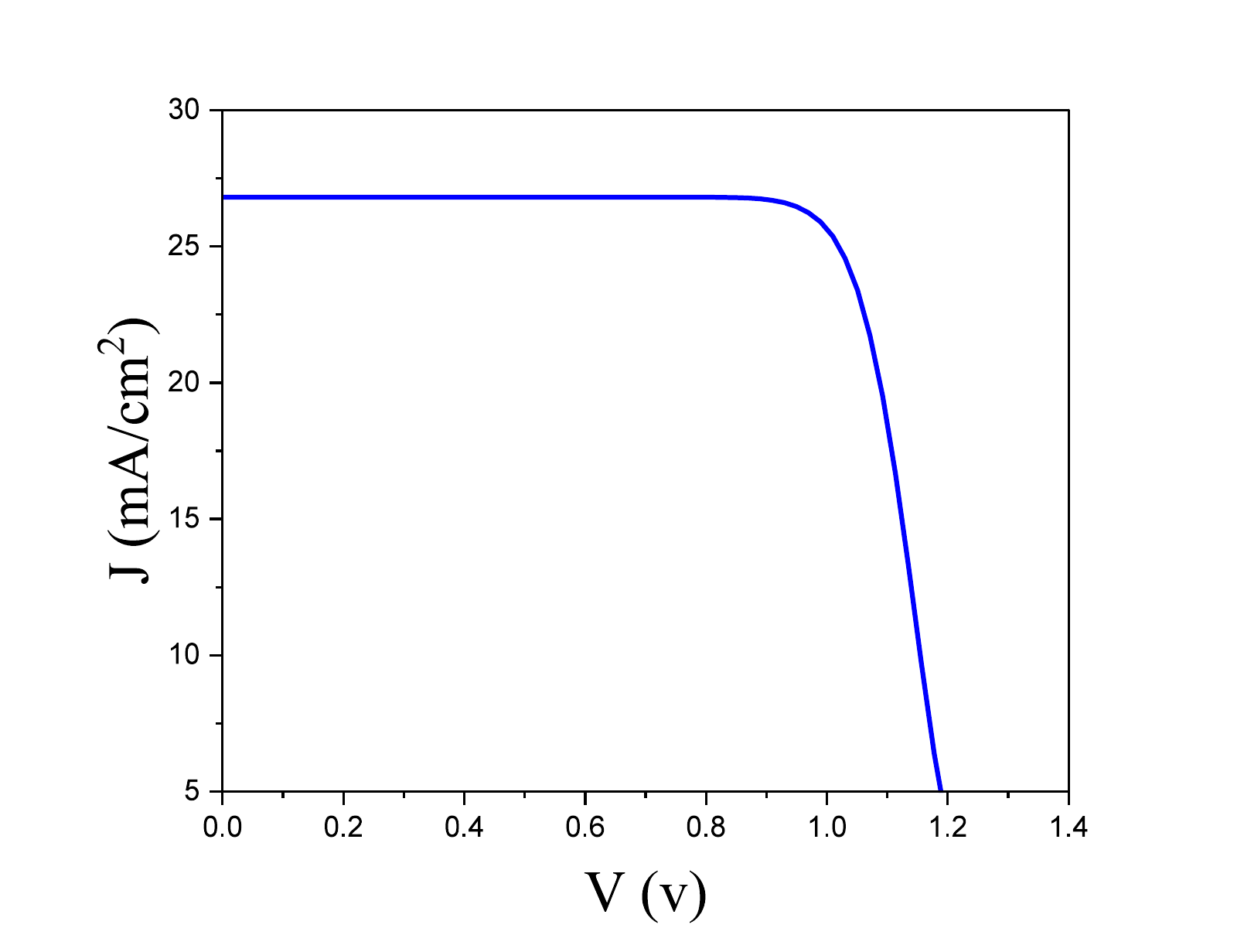}
    \caption{Current density–voltage (J‑V) characteristic of the FTO/ZnSe/CH$_3$NH$_3$PbI$_3$/NiO/Ag perovskite solar cell. Corresponding photovoltaic parameters are summarized in Table \ref{tab:top11}.}
    \label{fig:1}
\end{figure}

Fig \ref{fig:1} shows the J-V curve for the structure with the CH$_3$NH$_3$PbI$_3$ absorber layer. This common material has achieved high efficiencies in many device structures due to its favorable intrinsic properties. Its extremely high optical absorption coefficient enables even a thin layer to absorb a significant percentage of incident photons, an effect that is stronger in the visible spectrum. Additionally, the low effective mass and long charge carrier diffusion length reduce recombination, which ultimately enhances the PCE \cite{Oubihi2026-vd}.

A direct comparison of SCAPS-1D results across the three absorber materials reveals a clear inverse correlation between bandgap energy and device efficiency. This pattern is quantitatively corroborated by SHAP-based feature analysis, wherein the absorber bandgap is identified as one of the most influential parameters within the machine-learning  model. Consequently, both SCAPS-1D simulation and data-driven modeling converge on a unified design principle. Narrower bandgaps in the absorber layer critically enhance photovoltaic performance.

In addition to the absorber layer, other layers significantly influence the final efficiency of a perovskite solar cell, including the HTL. Among the 11 structures tested, NiO and MASnBr$_3$ exhibited the best performance out of the five candidate materials. This is partly attributed to their favorable band alignment with the absorber layer, which allows generated holes to reach the HTL without additional energy, thereby reducing recombination. Consistent with this, our machine learning analysis identified the HTL's band gap as the most influential parameter governing device efficiency. A wider band gap in this layer strongly suppresses interfacial recombination. Furthermore, a higher conduction band effective density of states in the HTL was found to facilitate rapid charge extraction. NiO also exhibits damp-heat stability, distinguishing it from other materials \cite{Shin2026-eg}. In addition to its suitable band alignment, MASnBr$_3$ possesses advantageous intrinsic properties, including its p‑type nature. In a p‑type material, holes serve as the majority carriers, thereby enabling more efficient hole extraction\cite{Fahad2026-eb}. From a modeling perspective, a lower hole mobility in the absorber layer appeared marginally beneficial likely by mitigating undesired carrier back‑transfer. These data‑driven insights quantitatively confirm that optimal HTL properties are as critical as the absorber itself for achieving high‑performance perovskite solar cells. 

With the exception of TiO$_2$, which has consistently shown reliable performance in previous studies, the selected ETL materials exhibited nearly similar behavior. TiO$_2$ offers high optical transparency, allowing maximum photon transmission to the absorber layer. Its hole-blocking property facilitates electron transport while suppressing holes. Moreover, the favorable energy band alignment of TiO$_2$ with the absorber layer further enhances its suitability for solar cell applications\cite{Hasnain2026-ar}. 

The efficiencies of the selected structures in this study range between 26$\%$ and 28$\%$ Achieving this high efficiency is directly related to proper photon absorption and the efficient conversion of incident photons into external electrical current. In the top-performing perovskite solar cells identified here, photon absorption typically reaches its maximum in the wavelength range of 400nm to 700nm, which corresponds to the visible light spectrum. As the wavelength of incident photons increases beyond 700 nm, the photon energy decreases. This reduction in energy may result in photons being unable to generate electron-hole pairs, in other words, the photon energy becomes smaller than the bandgap of the absorber layer. Consequently, beyond 700 nm, the photon absorption efficiency and its conversion into electrical energy generally decline. This trend is consistent with the lower performance observed for structures with indirect-bandgap absorbers such as Cs$_2$AgBiBr$_6$. On the other hand, at wavelengths below 400nm, the incident photons possess excessively high energy. This high energy can lead to rapid recombination of electrons and holes. This may offset the benefits of strong absorption and partially explain why even high-efficiency devices do not exceed 28$\%$ in this study.

\section{Conclusion}
In this work, we systematically investigated a combinatorial space of multilayer perovskite solar cells by combining SCAPS-1D simulations with machine learning and SHAP-based interpretability analysis. Using five ETL materials, five absorbers, and five HTLs, we evaluated 125 possible device architectures and identified several high-performance configurations with power conversion efficiencies exceeding 25$\%$. The best-performing structure, FTO/TiO$_2$/Cs$_2$AgBiBr$_6$/NiO/Ag, achieved a PCE of 28.85$\%$, demonstrating the strong potential of carefully engineered multilayer perovskite devices for high-efficiency photovoltaic applications.

The machine learning model, based on XGBoost and trained with a diverse representative subset of simulated devices, showed good predictive and ranking capability for screening unexplored configurations. The LOGO-CV results confirmed that the model could capture the overall relationship between material combinations and device performance, while SHAP analysis revealed the most influential physical descriptors governing efficiency. In particular, absorber bandgap, transport-layer band structure, and interfacial charge-extraction-related parameters were found to play key roles in determining photovoltaic performance.

Overall, the present study demonstrates that an integrated physics-based and data-driven workflow can accelerate the discovery of promising perovskite solar-cell architectures while reducing the need for exhaustive simulations. The results also provide physically meaningful guidance for future device design, emphasizing the importance of narrow bandgap absorbers, well-aligned transport layers, and optimized interfacial energetics. This framework can be extended to other photovoltaic material systems and may serve as a practical route toward efficient and computationally economical solar-cell optimization.

\printbibliography

@ARTICLE{Chen2017-ml,
  title     = "Thin single crystal perovskite solar cells to harvest
               below-bandgap light absorption",
  author    = "Chen, Zhaolai and Dong, Qingfeng and Liu, Ye and Bao, Chunxiong
               and Fang, Yanjun and Lin, Yun and Tang, Shi and Wang, Qi and
               Xiao, Xun and Bai, Yang and Deng, Yehao and Huang, Jinsong",
  journal   = "Nat. Commun.",
  publisher = "Springer Science and Business Media LLC",
  volume    =  8,
  number    =  1,
  pages     = "1890",
  month     =  dec,
  year      =  2017,
  language  = "en"
}

@ARTICLE{Zdanowicz2005-yo,
  title     = "Theoretical analysis of the optimum energy band gap of
               semiconductors for fabrication of solar cells for applications
               in higher latitudes locations",
  author    = "Zdanowicz, T and Rodziewicz, T and Zabkowska-Waclawek, M",
  journal   = "Sol. Energy Mater. Sol. Cells",
  publisher = "Elsevier BV",
  volume    =  87,
  number    = "1-4",
  pages     = "757--769",
  month     =  may,
  year      =  2005,
  language  = "en"
}

@ARTICLE{Hood2016-ee,
  title     = "Entropy and disorder enable charge separation in organic solar
               cells",
  author    = "Hood, Samantha N and Kassal, Ivan",
  journal   = "J. Phys. Chem. Lett.",
  publisher = "American Chemical Society (ACS)",
  volume    =  7,
  number    =  22,
  pages     = "4495--4500",
  month     =  nov,
  year      =  2016,
  language  = "en"
}

@ARTICLE{Klein2012-fz,
  title     = "Energy band alignment at interfaces of semiconducting oxides: A
               review of experimental determination using photoelectron
               spectroscopy and comparison with theoretical predictions by the
               electron affinity rule, charge neutrality levels, and the common
               anion rule",
  author    = "Klein, Andreas",
  journal   = "Thin Solid Films",
  publisher = "Elsevier BV",
  volume    =  520,
  number    =  10,
  pages     = "3721--3728",
  month     =  mar,
  year      =  2012,
  language  = "en"
}

@ARTICLE{Musiienko2024-nq,
  title     = "Resolving electron and hole transport properties in
               semiconductor materials by constant light-induced magneto
               transport",
  author    = "Musiienko, Artem and Yang, Fengjiu and Gries, Thomas William and
               Frasca, Chiara and Friedrich, Dennis and Al-Ashouri, Amran and
               Sa{\u g}lamkaya, Elifnaz and Lang, Felix and Kojda, Danny and
               Huang, Yi-Teng and Stacchini, Valerio and Hoye, Robert L Z and
               Ahmadi, Mahshid and Kanak, Andrii and Abate, Antonio",
  journal   = "Nat. Commun.",
  publisher = "Springer Science and Business Media LLC",
  volume    =  15,
  number    =  1,
  pages     = "316",
  month     =  jan,
  year      =  2024,
  copyright = "https://creativecommons.org/licenses/by/4.0",
  language  = "en"
}

@ARTICLE{Movla2023-tv,
  title     = "A numerical study on the relationship between the doping and
               performance in {P3HT:PCBM} organic bulk heterojunction solar
               cells",
  author    = "Movla, Hossein and Shahalizad, Afshin and Asgari, Asghar",
  journal   = "Sci. Rep.",
  publisher = "Springer Science and Business Media LLC",
  volume    =  13,
  number    =  1,
  pages     = "2031",
  month     =  feb,
  year      =  2023,
  copyright = "https://creativecommons.org/licenses/by/4.0",
  language  = "en"
}

@ARTICLE{Tsiligaridis2023-iq,
  title     = "Tree-based ensemble models, algorithms and performance measures
               for classification",
  author    = "Tsiligaridis, John",
  journal   = "Adv. Sci. Technol. Eng. Syst. J.",
  publisher = "ASTES Journal",
  volume    =  8,
  number    =  6,
  pages     = "19--25",
  month     =  nov,
  year      =  2023
}

@ARTICLE{Jiang2025-ne,
  title     = "Interpretable ensemble learning for materials property
               prediction with classical interatomic potentials",
  author    = "Jiang, Xinyu and Sun, Haofan and Choudhary, Kamal and Zhuang,
               Houlong and Nian, Qiong",
  journal   = "Npj Comput. Mater.",
  publisher = "Springer Science and Business Media LLC",
  volume    =  11,
  number    =  1,
  month     =  oct,
  year      =  2025,
  copyright = "https://creativecommons.org/licenses/by/4.0",
  language  = "en"
}

@ARTICLE{Ahamed2026-jp,
  title     = "Performance optimization and machine learning-guided parameter
               sensitivity analysis of lead-free {KGeCl3} perovskite solar
               cells",
  author    = "Ahamed, Tanzir and Bappy, Md Mehedi Hasan and Islam, Mohammad
               Rahimul and Uddin, Md Shihab and Hossain, Md Arafat and Ahammed,
               Tanvir",
  journal   = "RSC Adv.",
  publisher = "Royal Society of Chemistry (RSC)",
  volume    =  16,
  number    =  10,
  pages     = "8985--9011",
  month     =  feb,
  year      =  2026,
  copyright = "http://creativecommons.org/licenses/by/3.0/",
  language  = "en"
}

@ARTICLE{Bibi2025-kb,
  title     = "Machine learning-driven optimization of transport layer
               parameters in {CsSn} $_{0.5}$ Ge $_{0.5}$ {I} $_{3}$ perovskite
               solar cells",
  author    = "Bibi, Baseerat and Ur Rahman, Waseem and Us Sama, Najm and Guan,
               Linlin and Hatim Shah, Syed and Liu, Zhu",
  journal   = "IEEE Access",
  publisher = "Institute of Electrical and Electronics Engineers (IEEE)",
  volume    =  13,
  pages     = "185416--185432",
  year      =  2025,
  copyright = "https://creativecommons.org/licenses/by/4.0/legalcode"
}

@ARTICLE{De_la_Asuncion-Nadal2025-ze,
  title     = "Machine learning for perovskite solar cells: a comprehensive
               review on opportunities and challenges for materials scientists",
  author    = "de la Asunci{\'o}n-Nadal, V{\'\i}ctor and Iliffe Sprague,
               Christopher and Guijarro-Berdi{\~n}as, Bertha and Cappel, Ute B
               and Garc{\'\i}a-Fern{\'a}ndez, Alberto",
  journal   = "EES Solar",
  publisher = "Royal Society of Chemistry (RSC)",
  volume    =  1,
  number    =  6,
  pages     = "927--957",
  year      =  2025,
  copyright = "http://creativecommons.org/licenses/by-nc/3.0/",
  language  = "en"
}

@ARTICLE{Shwartz-Ziv2022-dy,
  title     = "Tabular data: Deep learning is not all you need",
  author    = "Shwartz-Ziv, Ravid and Armon, Amitai",
  journal   = "Inf. Fusion",
  publisher = "Elsevier BV",
  volume    =  81,
  pages     = "84--90",
  month     =  may,
  year      =  2022,
  language  = "en"
}

@ARTICLE{Abrar2025-cw,
  title     = "Optimization of Pb-free highly efficient {Cs$_2$AgInBr$_6$} double
               perovskite solar cells: A numerical investigation using {SCAPS}",
  author    = "Abrar, Mahir and Biswas, Ishrat Jahan and Hodges, Deidra",
  journal   = "J. Electron. Mater.",
  publisher = "Springer Science and Business Media LLC",
  volume    =  54,
  number    =  6,
  pages     = "4357--4365",
  month     =  jun,
  year      =  2025,
  copyright = "https://creativecommons.org/licenses/by/4.0",
  language  = "en"
}

@ARTICLE{Sangavi2025-en,
  title     = "Exploring {Sm2NiMnO6} as a lead-free absorber for perovskite
               solar cells: Insights from theoretical and experimental
               approaches",
  author    = "Sangavi, T and Vasanth, S and Viswanathan, C and Ponpandian, N",
  journal   = "Sol. Energy Mater. Sol. Cells",
  publisher = "Elsevier BV",
  volume    =  283,
  number    =  113456,
  pages     = "113456",
  month     =  may,
  year      =  2025,
  language  = "en"
}

@ARTICLE{Alkhammash2023-uu,
  title     = "Design and defect study of {Cs$_{2}$AgBiBr$_{6}$} double
               perovskite solar cell using suitable charge transport layers",
  author    = "Alkhammash, Hend I and Mottakin, M and Hossen, Md Mosaddek and
               Akhtaruzzaman, Md and Rashid, Mohammad Junaebur",
  journal   = "Semicond. Sci. Technol.",
  publisher = "IOP Publishing",
  volume    =  38,
  number    =  1,
  pages     = "015005",
  month     =  jan,
  year      =  2023,
  copyright = "http://creativecommons.org/licenses/by/4.0"
}

@ARTICLE{Uddin2024-cb,
  title     = "An in‐depth investigation of the combined optoelectronic and
               photovoltaic properties of lead‐free {cs$_{2}$AgBiBr$_{6}$}
               double perovskite solar cells using {DFT} and {SCAPS‐1D}
               frameworks",
  author    = "Uddin, M Shihab and Hossain, M Khalid and Uddin, Md Borhan and
               Toki, Gazi F I and Ouladsmane, Mohamed and Rubel, Mirza H K and
               Tishkevich, Daria I and Sasikumar, P and Haldhar, Rajesh and
               Pandey, Rahul",
  journal   = "Adv. Electron. Mater.",
  publisher = "Wiley",
  volume    =  10,
  number    =  5,
  month     =  may,
  year      =  2024,
  copyright = "http://creativecommons.org/licenses/by/4.0/",
  language  = "en"
}

@ARTICLE{Miah2024-rk,
  title     = "Band gap tuning of perovskite solar cells for enhancing the
               efficiency and stability: issues and prospects",
  author    = "Miah, Md Helal and Khandaker, Mayeen Uddin and Rahman, Md Bulu
               and Nur-E-Alam, Mohammad and Islam, Mohammad Aminul",
  journal   = "RSC Adv.",
  publisher = "Royal Society of Chemistry (RSC)",
  volume    =  14,
  number    =  23,
  pages     = "15876--15906",
  month     =  may,
  year      =  2024,
  copyright = "http://creativecommons.org/licenses/by-nc/3.0/",
  language  = "en"
}

@ARTICLE{Han2026-yd,
  title     = "{All-Inorganic} Cs $_{2}$ {AgInBr} $_{6}$ / {RbGeI} $_{3}$
               {PSCs} incorporating a p $^{+}$ {-MoS} $_{2}$ {TRL}",
  author    = "Han, Limei and Qu, Yingsu and Wu, Chenyu and Wang, Tao and
               Zhang, Meilin and Wu, Jiang",
  journal   = "J. Phys. Conf. Ser.",
  publisher = "IOP Publishing",
  volume    =  3171,
  number    =  1,
  pages     = "012011",
  month     =  jan,
  year      =  2026,
  copyright = "https://creativecommons.org/licenses/by/4.0/"
}

@ARTICLE{Gupta2026-zp,
  title     = "Unveiling pressure-driven transitions in {Cs$_2$AgBiBr$_6$}: Insights
               from {DFT} into a lead-free solar perovskite",
  author    = "Gupta, Sagita and Maurya, Devidutta and Srivastava, Sunil Kumar
               and Pareek, Umesh Kumar and Srivastava, Abhay P",
  
  journal   = "East Eur. J. Phys.",
  publisher = "V. N. Karazin Kharkiv National University",
  number    =  1,
  pages     = "363--372",
  month     =  mar,
  year      =  2026
}

@ARTICLE{Oubihi2026-vd,
  title     = "Methylammonium lead halide perovskites: Electronic structure and
               optical properties for tandem solar cells",
  author    = "Oubihi, Fatima and El Harouny, El Hassan and Ziani, Hanan and El
               Khamkhami, Jamal and Assaid, El Mahdi and Lachgar, Ahmed and
               Achahbar, Abdelfattah",
  journal   = "E3S Web Conf.",
  publisher = "EDP Sciences",
  volume    =  704,
  pages     = "01005",
  year      =  2026,
  copyright = "https://creativecommons.org/licenses/by/4.0/"
}

@ARTICLE{Shin2026-eg,
  title     = "Redox-active {NiOx-catalyzed} Li+ capture-extraction strategy
               for {tBP-free} {Spiro-OMeTAD} enables exceptional damp-heat
               stability in perovskite solar cells",
  author    = "Shin, Yun Seop and Kim, Minjin and Lee, Jaehwi and Yoon, Chang
               Hyeon and Seo, Jongdeuk and Choi, Gyeong-Cheon and Park, Sujung
               and Sung, Min Jung and Son, Kyungnan and Hong, Sungjun and
               Jeong, Inyoung and Byeon, Junseop and Jo, Yimhyun and Lee,
               Dongmin and Kim, Minseong and Cho, Shinuk and Seo, Ji-Youn and
               Kim, Jin Young and Kim, Dong Suk and Ahn, Sejin",
  journal   = "Adv. Sci. (Weinh.)",
  publisher = "Wiley",
  volume    =  13,
  number    =  25,
  pages     = "e21825",
  month     =  may,
  year      =  2026,
  copyright = "http://creativecommons.org/licenses/by/4.0/",
  language  = "en"
}

@ARTICLE{Fahad2026-eb,
  title     = "Device-level simulation of lead-free perovskite solar cells
               ({MASnI$_3$}, {MASnBr$_3$}, and {MABiI$_3$)}: A comparative study using
               {SCAPS-1D}",
  author    = "Fahad, M A A and Islam, M M and Mahmud, S and Narjim, S and
               Shanta, S and Khatun, M T and Alam, M M",
  journal   = "Next Energy",
  publisher = "Elsevier BV",
  volume    =  12,
  number    =  100645,
  pages     = "100645",
  month     =  jul,
  year      =  2026,
  copyright = "http://creativecommons.org/licenses/by-nc/4.0/",
  language  = "en"
}

@ARTICLE{Hasnain2026-ar,
  title     = "Performance evaluation of metal-doped {X-TiO2} electron
               transport layers in perovskite solar cell devices: a review",
  author    = "Hasnain, Syed M",
  journal   = "Bull. Mater. Sci. (India)",
  publisher = "Springer Science and Business Media LLC",
  volume    =  49,
  number    =  1,
  month     =  feb,
  year      =  2026,
  copyright = "https://www.springernature.com/gp/researchers/text-and-data-mining",
  language  = "en"
}

@ARTICLE{Saad2025-we,
  title     = "Role of solar energy in achieving global net-zero targets:
               Policy and technological perspective",
  author    = "Saad, Asadullah Muhammad Hossain and Myat, A",
  journal   = "Am. J. Energy Nat. Resour.",
  publisher = "E-palli",
  volume    =  4,
  number    =  1,
  pages     = "1--8",
  month     =  jan,
  year      =  2025
}

@ARTICLE{Li2026-zm,
  title     = "From lab to market: Strategies for stabilizing and scaling
               perovskite solar cells via printing technologies",
  author    = "Li, Xin and Aftab, Sikandar and Yewale, Manesh Ashok and Hegazy,
               Hosameldin Helmy and Akman, Erdi and Rubab, Najaf and Kus,
               Mahmut",
  journal   = "Energy Environ. Mater.",
  publisher = "Wiley",
  volume    =  9,
  number    =  1,
  month     =  jan,
  year      =  2026,
  copyright = "http://creativecommons.org/licenses/by/4.0/",
  language  = "en"
}

@ARTICLE{Kim2025-fp,
  title     = "Advanced interface engineering for perovskite solar cells: The
               way to ensure efficiency and stability",
  author    = "Kim, Dong Hoe and Park, Nam-Gyu",
  journal   = "Acc. Mater. Res.",
  publisher = "American Chemical Society (ACS)",
  volume    =  6,
  number    =  9,
  pages     = "1147--1157",
  month     =  sep,
  year      =  2025,
  language  = "en"
}

@ARTICLE{Ahmed2025-yt,
  title     = "Lead-free perovskites for solar cell applications: recent
               progress, ongoing challenges, and strategic approaches",
  author    = "Ahmed, Imtiaz and Prakash, Kamal and Mobin, Shaikh M",
  journal   = "Chem. Commun. (Camb.)",
  publisher = "Royal Society of Chemistry (RSC)",
  volume    =  61,
  number    =  37,
  pages     = "6691--6721",
  month     =  may,
  year      =  2025,
  language  = "en"
}

@ARTICLE{Biswas2026-zp,
  title     = "Investigating the optoelectronic properties and photovoltaic
               performance of {Na$_2$AuGaBr$_6$} based double perovskite solar cells
               via numerical simulation and {AI} techniques",
  author    = "Biswas, Bipul Chandra and Shimul, Asadul Islam and Paul,
               Indrojit and AlFaify, S and Benghanem, Mohamed and Rahman, Md
               Azizur and Solre, Gideon F B and Elboughdiri, Noureddine",
  journal   = "Sci. Rep.",
  publisher = "Springer Science and Business Media LLC",
  volume    =  16,
  number    =  1,
  month     =  feb,
  year      =  2026,
  copyright = "https://creativecommons.org/licenses/by/4.0",
  language  = "en"
}

@ARTICLE{Rangar2025-ao,
  title     = "Inorganic lead-free double perovskites for eco-friendly energy
               conversion technology",
  author    = "Rangar, Kailash and Rawat, Kanchan and Tariyal, Shivanika and
               Sharma, Kamal Nayan and Kumar, Kishor and Ahuja, Ushma and Soni,
               Amit and Sahariya, Jagrati",
  journal   = "J. Mater. Sci.: Mater. Electron.",
  publisher = "Springer Science and Business Media LLC",
  volume    =  36,
  number    =  35,
  month     =  dec,
  year      =  2025,
  copyright = "https://www.springernature.com/gp/researchers/text-and-data-mining",
  language  = "en"
}

@ARTICLE{Hossain2026-xv,
  title     = "Performance boost in novel {CBAI/SNMO} double perovskite solar
               cells via {GO-induced} back surface field optimization",
  author    = "Hossain, Md Faruk and Rahman, Md Mahabur and Mia, Md Masum and
               Alharbi, Abdullah Marzouq and Badi, Nacer and Elboughdiri,
               Noureddine and Amami, Mongi and Ben Farhat, Lamia and Rahman, Md
               Ferdous",
  journal   = "Phys. Chem. Chem. Phys.",
  publisher = "Royal Society of Chemistry (RSC)",
  volume    =  28,
  number    =  19,
  pages     = "11795--11815",
  month     =  may,
  year      =  2026,
  language  = "en"
}

@ARTICLE{Fatmi2025-ec,
  title    = "High performance double perovskites of {Cs$_2$InAgBr$_6$} and
              {Cs$_2$InAgCl$_6$} structural electronic optical and thermoelectric
              properties for next generation photovoltaics",
  author   = "Fatmi, M and Bouferrache, K and Ghebouli, M A and Ghebouli, B and
              Alanazi, Faisal Katib and Albaqami, Munirah D and Mohammad, Saikh
              and Benali, A",
  journal  = "Sci. Rep.",
  volume   =  15,
  number   =  1,
  pages    = "20851",
  month    =  jul,
  year     =  2025,
  language = "en"
}

@article{moses,
author = {Irungu, Moses},
year = {2025},
month = {12},
pages = {100},
title = {Comparative First-Principles Study of Lead-Free Cs$_2$AgB$'$Br$_6$ (B$'$ = Bi, Sb, In) Double Perovskites for Photovoltaic Applications},
volume = {12},
journal = {International Journal of Emerging Research in Engineering and Technology}
}

@ARTICLE{Borah2026-fq,
  title     = "Design and optimization of {BaSnO3/SW-CNT} hybrid electron
               extraction architecture enabling extended spectral response in
               nontoxic {Cs$_2$AgBiBr$_6$} double perovskite photovoltaic cells:
               Insights from {SCAPS-1D}",
  author    = "Borah, Janmoni and Baruah, Smriti",
  journal   = "Physica B Condens. Matter",
  publisher = "Elsevier BV",
  volume    =  724,
  number    =  418178,
  pages     = "418178",
  month     =  feb,
  year      =  2026,
  language  = "en"
}

@ARTICLE{Li2022-mb,
  title     = "Pinning bromide ion with ionic liquid in lead‐free
               {cs$_{2}$AgBiBr$_{6}$} double perovskite solar cells",
  author    = "Li, Jiangning and Meng, Xianghuan and Wu, Zhiheng and Duan,
               Yanyan and Guo, Ruxin and Xiao, Weidong and Zhang, Yongshang and
               Li, Yukun and Shen, Yonglong and Zhang, Wei and Shao, Guosheng",
  journal   = "Adv. Funct. Mater.",
  publisher = "Wiley",
  volume    =  32,
  number    =  25,
  pages     = "2112991",
  month     =  jun,
  year      =  2022,
  copyright = "http://onlinelibrary.wiley.com/termsAndConditions\#vor",
  language  = "en"
}

@ARTICLE{Pang2022-py,
  title     = "Improved charge extraction and atmospheric stability of
               all-inorganic {Cs$_2$AgBiBr$_6$} perovskite solar cells by {MoS2}
               nanoflakes",
  author    = "Pang, Beili and Chen, Xiang and Bao, Fujie and Liu, Yili and
               Feng, Ting and Dong, Hongzhou and Yu, Liyan and Dong, Lifeng",
  journal   = "Sol. Energy Mater. Sol. Cells",
  publisher = "Elsevier BV",
  volume    =  246,
  number    =  111932,
  pages     = "111932",
  month     =  oct,
  year      =  2022,
  language  = "en"
}

@ARTICLE{Ou2022-cl,
  title     = "Boosting the stability and efficiency of {Cs$_2$AgBiBr$_6$} perovskite
               solar cells via Zn doping",
  author    = "Ou, Yingjun and Lu, Zuizhi and Lu, Jiangying and Zhong, Xiaoying
               and Chen, Peican and Zhou, Liya and Chen, Ting",
  journal   = "Opt. Mater. (Amst.)",
  publisher = "Elsevier BV",
  volume    =  129,
  number    =  112452,
  pages     = "112452",
  month     =  jul,
  year      =  2022,
  language  = "en"
}

@INPROCEEDINGS{Singh2026-pe,
  title           = "Predictive modelling of perovskite solar cell structure
                     using machine learning: Process optimization and device
                     performance",
  booktitle       = "2026 International Conference on Electric Power and
                     Renewable Energy ({EPREC})",
  author          = "Singh, Deepak Kumar and Singh, Jyoti Prashant and Patel,
                     Alok Kumar and Gupta, Saurabh and Yadava, Prem Chand and
                     Yadav, Vishal",
  publisher       = "IEEE",
  pages           = "1--6",
  month           =  jan,
  year            =  2026,
  conference      = "2026 International Conference on Electric Power and
                     Renewable Energy (EPREC)",
  location        = "Durg, India"
}

@ARTICLE{Boretti2026-dx,
  title     = "Metal chalcogenide nanostructures: A bifunctional platform to
               resolve the efficiency-stability trade-off in commercial
               perovskite solar cells",
  author    = "Boretti, Alberto",
  journal   = "Next Mater.",
  publisher = "Elsevier BV",
  volume    =  10,
  number    =  101580,
  pages     = "101580",
  month     =  jan,
  year      =  2026,
  copyright = "http://creativecommons.org/licenses/by-nc-nd/4.0/",
  language  = "en"
}

@ARTICLE{McGovern2021-ct,
  title     = "Grain size influences activation energy and migration pathways
               in {MAPbBr$_3$} perovskite solar cells",
  author    = "McGovern, Lucie and Koschany, Isabel and Grimaldi, Gianluca and
               Muscarella, Loreta A and Ehrler, Bruno",
  journal   = "J. Phys. Chem. Lett.",
  publisher = "American Chemical Society (ACS)",
  volume    =  12,
  number    =  9,
  pages     = "2423--2428",
  month     =  mar,
  year      =  2021,
  copyright = "https://creativecommons.org/licenses/by-nc-nd/4.0/",
  language  = "en"
}

@ARTICLE{Chen2026-va,
  title     = "Simulation and optimization of {MAPbI} $_{2}$ Br and {MAPbBr}
               $_{3}$ perovskite solar cells achieving efficiencies up to
               27.9\%",
  author    = "Chen, Yaolan and Zhao, Feng and Ji, Kaixuan and Feng, Zisheng
               and Chen, Leyi and Zhao, Fangying and Deng, Quanrong and Jiang,
               Tingwei and Liu, Ziyuan",
  journal   = "Phys. Scr.",
  publisher = "IOP Publishing",
  volume    =  101,
  number    =  16,
  pages     = "165507",
  month     =  apr,
  year      =  2026,
  copyright = "https://publishingsupport.iopscience.iop.org/iop-standard/v1"
}

@ARTICLE{Mirdoraghi2025-fp,
  title     = "Energy band engineering and cubic crystallinity: achieving 36\%
               efficiency in lead-free {MASnBr3} perovskite solar cells via
               {SCAPS-guided} optimization",
  author    = "Mirdoraghi, Mohammad and Shakiba, Maryam and Khademalrasool,
               Marzieh",
  journal   = "Opt. Quantum Electron.",
  publisher = "Springer Science and Business Media LLC",
  volume    =  57,
  number    =  6,
  month     =  jun,
  year      =  2025,
  copyright = "https://www.springernature.com/gp/researchers/text-and-data-mining",
  language  = "en"
}

@ARTICLE{Hasan2025-te,
  title     = "Next-generation lead-free solar cells with {MASnBr3/ZnSnN2} dual
               absorbers for high efficiency",
  author    = "Hasan, Md Mehedi and Siddika, Mst Aysha and Ali, Md Feroz and
               Sheikh, Md Rafiqul Islam and Al Mamun, Abdullah and Hossen, Md
               Jakir",
  journal   = "Front. Mater.",
  publisher = "Frontiers Media SA",
  volume    =  12,
  number    =  1652733,
  month     =  aug,
  year      =  2025,
  copyright = "https://creativecommons.org/licenses/by/4.0/"
}

@ARTICLE{Dakua2025-lx,
  title     = "Study on the feasibility of high {PCE} {C} $_{2}$ {N} solar cell
               incorporating a novel {HT} material",
  author    = "Dakua, Pratap Kumar and Sri, Karri Hema and Preetham, K and
               Nutan, M and Shaik, K Baba",
  journal   = "ChemistrySelect",
  publisher = "Wiley",
  volume    =  10,
  number    =  21,
  month     =  jun,
  year      =  2025,
  copyright = "http://onlinelibrary.wiley.com/termsAndConditions\#vor",
  language  = "en"
}

@ARTICLE{Singh2026-nw,
  title     = "High-efficiency lead-free {MASnI3} perovskite solar cells using
               {ZnSe} {ETL} and {NiO} {HTL}: Optimization and comparative study",
  author    = "Singh, Deepak Kumar and Patel, Alok Kumar and Srivastava,
               Vaibhava and Mishra, Rajan and Soni, Sanjay Kumar",
  journal   = "J. Phys. Chem. Solids",
  publisher = "Elsevier BV",
  volume    =  211,
  number    =  113507,
  pages     = "113507",
  month     =  apr,
  year      =  2026,
  language  = "en"
}

@ARTICLE{Lotfy2025-lc,
  title     = "Numerical simulation and optimization of {FTO/TiO2/CZTS/CuO/Au}
               solar cell using {SCAPS-1D}",
  author    = "Lotfy, Lofty A and Abdelfatah, Mahmoud and Sharshir, Swellam W
               and El-Naggar, Ahmed A and Ismail, Walid and El-Shaer,
               Abdelhamid",
  journal   = "Sci. Rep.",
  publisher = "Springer Science and Business Media LLC",
  volume    =  15,
  number    =  1,
  pages     = "28022",
  month     =  jul,
  year      =  2025,
  copyright = "https://creativecommons.org/licenses/by/4.0",
  language  = "en"
}

@ARTICLE{Roula2025-tz,
  title     = "Optimized lead-free perovskite solar cell with {SnS$_2$} {ETL} and
               {Spiro-OMeTAD} {HTL} for > 33\% efficiency",
  author    = "Roula, Srinivash and Jyothula, Hari and Dakua, Pratap Kumar and
               Pattanayak, Prabina",
  journal   = "Discov Electron",
  publisher = "Springer Science and Business Media LLC",
  volume    =  2,
  number    =  1,
  month     =  sep,
  year      =  2025,
  copyright = "https://creativecommons.org/licenses/by-nc-nd/4.0",
  language  = "en"
}

@ARTICLE{Khan2025-tu,
  title     = "Numerical simulation and performance enhancement of
               {CsBi3I10-based} heterojunction solar cell with various
               semiconductor layers ({CZTS}, {CZTGS}, {Al0.8Ga0.2Sb}, {GaAs})
               along with machine learning-based analysis",
  author    = "Khan, Rabeya and Farjana, Nadira and Akter Jim, Mst Jahida and
               Al-Humaidi, Jehan Y and Islam, Md Rasidul and Rana, Md Masud",
  journal   = "Sol. Energy",
  publisher = "Elsevier BV",
  volume    =  295,
  number    =  113539,
  pages     = "113539",
  month     =  jul,
  year      =  2025,
  language  = "en"
}

@ARTICLE{Monga2025-lj,
  title     = "Ag doped {ZnSe} as electron transport layer with enhanced
               electron mobility and better band alignment for efficient lead
               free - all inorganic {CuBiSC$_{l2}$} perovskite solar cells: a
               simulation",
  author    = "Monga, Kamil and Singh, Vasundhara and Parmar, Avanish Singh and
               Chaudhary, Shilpi",
  journal   = "Phys. Scr.",
  publisher = "IOP Publishing",
  volume    =  100,
  number    =  8,
  pages     = "085948",
  month     =  aug,
  year      =  2025,
  copyright = "https://iopscience.iop.org/page/copyright"
}

@ARTICLE{Shafique2026-fw,
  title     = "Performance enhancement of {HTL} free perovskite solar cells
               through {ETL} and back contact engineering",
  author    = "Shafique, Amina and Amin, Uzma and Abu-Siada, Ahmed",
  journal   = "J. Solgel Sci. Technol.",
  publisher = "Springer Science and Business Media LLC",
  volume    =  117,
  number    =  3,
  month     =  feb,
  year      =  2026,
  copyright = "https://creativecommons.org/licenses/by/4.0",
  language  = "en"
}

@ARTICLE{Saidani2025-vr,
  title    = "Probing high-efficiency
              {Cs0.05(FA0.77MA0.23)0.95Pb(I0.77Br0.23)3-based} perovskite solar
              cells through first principles computations and {SCAPS-1D}
              simulation",
  author   = "Saidani, Okba and Goumri-Said, Souraya and Yousfi, Abderrahim and
              Sahoo, Girija Shankar and Kanoun, Mohammed Benali",
  journal  = "RSC Adv.",
  volume   =  15,
  number   =  10,
  pages    = "7342--7353",
  month    =  mar,
  year     =  2025,
  language = "en"
}

@ARTICLE{Zhou2025-wb,
  title     = "A chain entanglement gelled {SnO$_2$} electron transport layer for
               enhanced perovskite solar cell performance and effective lead
               capture",
  author    = "Zhou, Yuchen and He, Zhengyan and Wei, Qilin and Sun, Anni and
               Wu, Zilong and Huang, Dan and Zhang, Shufang and Yu, William W",
  journal   = "Adv. Mater.",
  publisher = "Wiley",
  volume    =  37,
  number    =  8,
  pages     = "e2416932",
  month     =  feb,
  year      =  2025,
  copyright = "http://onlinelibrary.wiley.com/termsAndConditions\#vor",
  language  = "en"
}

@ARTICLE{Talukder2025-uh,
  title     = "Performance optimization of Cs $_{2}$ {PtBr} $_{6}$ -based
               perovskite solar cells through {ETL} and {HTL} engineering: a
               {SCAPS-1D} simulation study",
  author    = "Talukder, Md Sourav and Mahmud, Sultan and Rana, Md Masud and
               Rabbi, M D Shahadat Khan and Islam, Mohammad Rahimul and
               Rukonuzzaman, M",
  journal   = "Phys. Scr.",
  publisher = "IOP Publishing",
  volume    =  100,
  number    =  12,
  pages     = "125970",
  month     =  dec,
  year      =  2025,
  copyright = "https://publishingsupport.iopscience.iop.org/iop-standard/v1"
}

@ARTICLE{Mastoor2026-jo,
  title     = "Scalable machine learning models for predicting quantum
               transport in disordered {2D} hexagonal materials",
  author    = "Mastoor, Seyed Mahdi and Baghervand, Ayda and Kordbacheh,
               Amirhossein Ahmadkhan",
  journal   = "Comput. Mater. Sci.",
  publisher = "Elsevier BV",
  volume    =  266,
  number    =  114561,
  pages     = "114561",
  month     =  feb,
  year      =  2026,
  language  = "en"
}

@ARTICLE{Peivaste2025-hj,
  title     = "Artificial intelligence in materials science and engineering:
               Current landscape, key challenges, and future trajectories",
  author    = "Peivaste, Iman and Belouettar, Salim and Mercuri, Francesco and
               Fantuzzi, Nicholas and Dehghani, Hamidreza and Izadi, Razie and
               Ibrahim, Halliru and Lengiewicz, Jakub and Belouettar-Mathis,
               Ma{\"e}l and Bendine, Kouider and Makradi, Ahmed and Horsch,
               Martin and Klein, Peter and Hachemi, Mohamed El and Preisig,
               Heinz A and Rezgui, Yacine and Konchakova, Natalia and Daouadji,
               Ali",
  journal   = "Compos. Struct.",
  publisher = "Elsevier BV",
  volume    =  372,
  number    =  119419,
  pages     = "119419",
  month     =  nov,
  year      =  2025,
  language  = "en"
}

@ARTICLE{Mao2025-oh,
  title     = "A comprehensive review of machine learning applications in
               perovskite solar cells: Materials discovery, device performance,
               process optimization and systems integration",
  author    = "Mao, Ling and Xiang, Changying",
  journal   = "Mater. Today Energy",
  publisher = "Elsevier BV",
  volume    =  47,
  number    =  101742,
  pages     = "101742",
  month     =  jan,
  year      =  2025,
  language  = "en"
}

@ARTICLE{Shafian2025-xs,
  title     = "Predicting high-performance perovskite solar cells using
               {AI-based} machine learning models",
  author    = "Shafian, Shafidah and Husen, Mohd Nizam and Xie, Lin and Kim,
               Kyungkon",
  journal   = "Materials Today Sustainability",
  publisher = "Elsevier BV",
  volume    =  31,
  number    =  101176,
  pages     = "101176",
  month     =  sep,
  year      =  2025,
  copyright = "http://creativecommons.org/licenses/by-nc-nd/4.0/",
  language  = "en"
}

@ARTICLE{Fahim2026-fw,
  title     = "Simulation, optimization, and machine learning strategies for
               {CH$_3$NH$_3$PbBr$_3$} perovskite solar cells",
  author    = "Fahim, Safikur Rahman and Sarker, Md Shamim and Piya, Mahzabin
               Islam and Bhuiyan, Jubaer Ahamed and Mamur, Hayati and Bhuiyan,
               Mohammad Ruhul Amin",
  journal   = "Next Energy",
  publisher = "Elsevier BV",
  volume    =  10,
  number    =  100491,
  pages     = "100491",
  month     =  jan,
  year      =  2026,
  copyright = "http://creativecommons.org/licenses/by-nc-nd/4.0/",
  language  = "en"
}

@ARTICLE{Bechane2025-sr,
  title     = "Numerical simulations for the performance optimization of
               {SnO$_2$/Cs$_2$AgInBr$_6$/CuO} lead-free perovskite solar cells",
  author    = "Bechane, Leila and Benguesmia, Hani and Hadda, Tiouiri",
  journal   = "Eng. Technol. Appl. Sci. Res.",
  publisher = "Engineering, Technology \& Applied Science Research",
  volume    =  15,
  number    =  6,
  pages     = "28706--28709",
  month     =  dec,
  year      =  2025
}

@ARTICLE{Sevillano-Bendezu2023-kv,
  title     = "Predictability and interrelations of spectral indicators for
               {PV} performance in multiple latitudes and climates",
  author    = "Sevillano-Bendez{\'u}, M A and Khenkin, M and Nofuentes, G and
               de la Casa, J and Ulbrich, C and T{\"o}fflinger, J A",
  journal   = "Sol. Energy",
  publisher = "Elsevier BV",
  volume    =  259,
  pages     = "174--187",
  month     =  jul,
  year      =  2023,
  copyright = "http://creativecommons.org/licenses/by/4.0/",
  language  = "en"
}

@ARTICLE{Taheri2021-nr,
  title     = "Effect of defects on high efficient perovskite solar cells",
  author    = "Taheri, Sara and Ahmadkhan kordbacheh, Amirhossein and Minbashi,
               Mehran and Hajjiah, Ali",
  journal   = "Opt. Mater. (Amst.)",
  publisher = "Elsevier BV",
  volume    =  111,
  number    =  110601,
  pages     = "110601",
  month     =  jan,
  year      =  2021,
  language  = "en"
}

\end{document}